\documentclass{bmcart}

\usepackage[utf8]{inputenc} %unicode support
\usepackage{graphicx}
\startlocaldefs
\endlocaldefs

\begin{document}

%%% Start of article front matter
\begin{frontmatter}

\begin{fmbox}
\dochead{Review}

%%%%%%%%%%%%%%%%%%%%%%%%%%%%%%%%%%%%%%%%%%%%%%
%%                                          %%
%% Enter the title of your article here     %%
%%                                          %%
%%%%%%%%%%%%%%%%%%%%%%%%%%%%%%%%%%%%%%%%%%%%%%

\title{Epidemics, the Ising-model and percolation theory: a comprehensive review focussed on Covid-19}

%%%%%%%%%%%%%%%%%%%%%%%%%%%%%%%%%%%%%%%%%%%%%%
%%                                          %%
%% Enter the authors here                   %%
%%                                          %%
%% Specify information, if available,       %%
%% in the form:                             %%
%%   <key>={<id1>,<id2>}                    %%
%%   <key>=                                 %%
%% Comment or delete the keys which are     %%
%% not used. Repeat \author command as much %%
%% as required.                             %%
%%                                          %%
%%%%%%%%%%%%%%%%%%%%%%%%%%%%%%%%%%%%%%%%%%%%%%

\author[
   addressref={aff1},                   % id's of addresses, e.g. {aff1,aff2}
   %corref={aff1},                       % id of corresponding address, if any
   %noteref={n1},                        % id's of article notes, if any
   %email={}   % email address
]{\inits{IFM}\fnm{Isys F.} \snm{Mello}}
\author[
   addressref={aff1},                   % id's of addresses, e.g. {aff1,aff2}
   %corref={aff1},                       % id of corresponding address, if any
   %noteref={n1},                        % id's of article notes, if any
   %email={}   % email address
]{\inits{LS}\fnm{Lucas} \snm{Squillante}}
\author[
   addressref={aff2,aff3},                   % id's of addresses, e.g. {aff1,aff2}
   %corref={aff1},                       % id of corresponding address, if any
   %noteref={n1},                        % id's of article notes, if any
   %email={}   % email address
]{\inits{GOG}\fnm{Gabriel O.} \snm{Gomes}}
\author[
   addressref={aff4},                   % id's of addresses, e.g. {aff1,aff2}
   %corref={aff1},                       % id of corresponding address, if any
   %noteref={n1},                        % id's of article notes, if any
   %email={}   % email address
]{\inits{ACS}\fnm{Antonio C.} \snm{Seridonio}}
\author[
   addressref={aff1},                   % id's of addresses, e.g. {aff1,aff2}
   %corref={aff1},                       % id of corresponding address, if any
   %noteref={n1},                        % id's of article notes, if any
   email={mariano.souza@unesp.br}   % email address
]{\inits{MDS}\fnm{Mariano} \snm{de Souza}}

%%%%%%%%%%%%%%%%%%%%%%%%%%%%%%%%%%%%%%%%%%%%%%
%%                                          %%
%% Enter the authors' addresses here        %%
%%                                          %%
%% Repeat \address commands as much as      %%
%% required.                                %%
%%                                          %%
%%%%%%%%%%%%%%%%%%%%%%%%%%%%%%%%%%%%%%%%%%%%%%

\address[id=aff1]{%                           % unique id
  \orgname{S\~ao Paulo State University (Unesp), IGCE - Physics Department, Rio Claro - SP, Brazil} % university, etc
  \street{}                     %
  \postcode{}                                % post or zip code
  %\city{}                              % city
  %\cny{}                                    % country
}
\address[id=aff2]{%
  \orgname{University of S\~ao Paulo, Department of Astronomy, SP, Brazil}
  %\street{}
  %\postcode{}
  %\city{}
  %\cny{}
}
\address[id=aff3]{%
  \orgname{Present address: Observatoire de Gen\`eve, Universit\'e de Gen\`eve, 51 Chemin des Maillettes, CH-1290 Sauverny, Switzerland}
  %\street{}
  %\postcode{}
  %\city{}
  %\cny{}
}
\address[id=aff4]{%
  \orgname{S\~ao Paulo State University (Unesp), Department of Physics and Chemistry, Ilha Solteira - SP, Brazil}
  %\street{}
  %\postcode{}
  %\city{}
  %\cny{}
}

%%%%%%%%%%%%%%%%%%%%%%%%%%%%%%%%%%%%%%%%%%%%%%
%%                                          %%
%% Enter short notes here                   %%
%%                                          %%
%% Short notes will be after addresses      %%
%% on first page.                           %%
%%                                          %%
%%%%%%%%%%%%%%%%%%%%%%%%%%%%%%%%%%%%%%%%%%%%%%

%\begin{artnotes}
%\note{Sample of title note}     % note to the article
%\note[id=n1]{Equal contributor} % note, connected to author
%\end{artnotes}

\end{fmbox}% comment this for two column layout

%%%%%%%%%%%%%%%%%%%%%%%%%%%%%%%%%%%%%%%%%%%%%%
%%                                          %%
%% The Abstract begins here                 %%
%%                                          %%
%% Please refer to the Instructions for     %%
%% authors on http://www.biomedcentral.com  %%
%% and include the section headings         %%
%% accordingly for your article type.       %%
%%                                          %%
%%%%%%%%%%%%%%%%%%%%%%%%%%%%%%%%%%%%%%%%%%%%%%

\begin{abstractbox}

\begin{abstract} % abstract
The recent spread of Covid-19 (Coronavirus disease) all over the world has made it one of the most important and discussed topic nowadays. Since its outbreak in Wuhan, several investigations have been carried out aiming to describe and support the containment of the disease spread. Here, we revisit well-established concepts of epidemiology, the Ising-model, and  percolation theory. Also, we employ a spin $S$ = 1/2 Ising-like model and a (logistic) Fermi-Dirac-like function to describe the spread of Covid-19. Our analysis reinforces well-established literature results, namely: \emph{i}) that the epidemic curves can be described by a Gaussian-type function; \emph{ii}) that the temporal evolution of the accumulative number of infections and fatalities follow a logistic function, which has some resemblance with a distorted Fermi-Dirac-like function; \emph{iii}) the key role played by the quarantine to block the spread of Covid-19 in terms of an \emph{interacting} parameter, which emulates the contact between infected and non-infected people. Furthermore, in the frame of elementary percolation theory, we show that: \emph{i}) the percolation probability can be associated with the probability of a person being infected with Covid-19; \emph{ii}) the concepts of blocked and non-blocked connections can be associated, respectively, with a person respecting or not the social distancing, impacting thus in the probability of an infected person to infect other people. Increasing the number of infected people leads to an increase in the number of net connections, giving rise thus to a higher probability of new infections (percolation). We demonstrate the importance of social distancing in preventing the spread of Covid-19 in a pedagogical way.  Given the impossibility of making a precise forecast of the disease spread, we highlight the importance of taking into account additional factors, such as climate changes and urbanization, in the mathematical description of epidemics. Yet, we make a connection between the standard mathematical models employed in epidemics and well-established concepts in condensed matter Physics, such as the Fermi gas and the Landau Fermi-liquid picture.
\end{abstract}

%%%%%%%%%%%%%%%%%%%%%%%%%%%%%%%%%%%%%%%%%%%%%%
%%                                          %%
%% The keywords begin here                  %%
%%                                          %%
%% Put each keyword in separate \kwd{}.     %%
%%                                          %%
%%%%%%%%%%%%%%%%%%%%%%%%%%%%%%%%%%%%%%%%%%%%%%

\begin{keyword}
\kwd{Covid-19}
\kwd{percolation theory}
\kwd{Ising-model}
\kwd{logistic function.}
\end{keyword}

% MSC classifications codes, if any
%\begin{keyword}[class=AMS]
%\kwd[Primary ]{}
%\kwd{}
%\kwd[; secondary ]{}
%\end{keyword}

\end{abstractbox}
%
%\end{fmbox}% uncomment this for twcolumn layout

\end{frontmatter}

%%%%%%%%%%%%%%%%%%%%%%%%%%%%%%%%%%%%%%%%%%%%%%
%%                                          %%
%% The Main Body begins here                %%
%%                                          %%
%% Please refer to the instructions for     %%
%% authors on:                              %%
%% http://www.biomedcentral.com/info/authors%%
%% and include the section headings         %%
%% accordingly for your article type.       %%
%%                                          %%
%% See the Results and Discussion section   %%
%% for details on how to create sub-sections%%
%%                                          %%
%% use \cite{...} to cite references        %%
%%  \cite{koon} and                         %%
%%  \cite{oreg,khar,zvai,xjon,schn,pond}    %%
%%  \nocite{smith,marg,hunn,advi,koha,mouse}%%
%%                                          %%
%%%%%%%%%%%%%%%%%%%%%%%%%%%%%%%%%%%%%%%%%%%%%%

%%%%%%%%%%%%%%%%%%%%%%%%% start of article main body
% <put your article body there>

%%%%%%%%%%%%%%%%
%% Background %%
%%
\section{Introduction}
 In the field of epidemics, every hour counts and there is an urge in predicting the temporal evolution of the disease aiming to find the best way to deal with it and to establish  a proper control of its spread. Recently, the pandemic Covid-19 (Coronavirus disease) has been rapidly spreading all over the world, being needless to mention the impact of it in our lives in a broad context, see, e.g., Refs.\,\cite{Chinazzi,Servick,Cohen,Rzymski,Liu,Vinko,Squazzoni2020}. It has been proposed that such a quick spread of Covid-19  in human beings is associated with a spike protein, which in turn has a site that is triggered by an enzyme called furin. The latter lies dangling on the surface of the virus, leading thus unfortunately to the infection of human cells much more easily \cite{Mallapaty}. Therefore, there is an urgency of an appropriate mathematical description of the spread of Covid-19 aiming to contain the disease. This is particularly true aiming to support the health government agencies all over the world to maximize the effectiveness of medical support strategy in such a global crisis.

\subsection{The SIR model}
 In the field of epidemiology \cite{Foppa, Kramer, Giesecke, White,Shlomo}, several mathematical approaches aiming to describe the infectious disease spread have been employed. Among them, the SIR (Susceptible, Infectious, Removed) model \cite{Giesecke,Nelson2014,Frauenthal2012,Ma2009,Brauer2008} stands out. Such a model considers the contact transmission risk, the average number of contacts between people as a function of time $t$, and the time that a person remains infected and thus can infect others.  The latter enable us to infer the average number of people $R_0$ directly infected. The value of $R_0$ usually dictates if the disease will eventually disappear ($R_0 < 1$), if an endemic will take place ($R_0 = 1$) or even if there will be an epidemic ($R_0 > 1$) \cite{Giesecke}. Furthermore, three variables provide key information about the disease, namely the fraction of people $S$ that are susceptible of being infected, the fraction of people $I'$ that are already infected and can thus infect others, and the fraction of people $R$ that becomes immune to the disease. The epidemic curve is achieved employing the following  three coupled first-order
differential equations \cite{Giesecke}:
\begin{eqnarray}
\frac{dS}{dt}&=&-\beta'\kappa S I'; \label{susceptible} \\
\frac{dI'}{dt}&=&\beta'\kappa S I' - \frac{I'}{D}; \label{infected} \\
\frac{dR}{dt}&=&\frac{I'}{D}, \label{removed}
\end{eqnarray}
where $\beta'$ is the contact transmission risk, $\kappa$ is the average number of contacts between people as a function of time, and $D$ refers to the time that a person remains infected and, as a consequence, can infect others. Also, using the previously defined parameters $R_0$ can be written as $R_0 = \beta'\kappa D$. Although the system of differential equations defined in the frame of the SIR model describes nicely the behavior of epidemics, the $\beta'$, $\kappa$, and consequently, $R_0$ factors are considered constant. However, such factors can be changed over time influenced by other parameters such as social distancing, for instance. Thus, a time dependence of $\beta'$ and $\kappa$ has to be taken into account, cf.\,discussion in Refs.\,\cite{Chen2020,Dehning2020}.
Given the non-linearity of the system of equations, the mathematical solution is non-trivial, being numerical analysis required in many cases. Now, for the sake of completeness, we will analyze each differential equation separately. The negative sign in Eq.\,\ref{susceptible} indicates that $S(t)$ decreases monotonically with time. Note that the factor $\beta'\kappa S I$ is also present in Eq.\,\ref{infected}, but with opposite sign, i.e.,
 as the number of infected people increases, less people are indeed susceptible, since a fraction of the susceptible people has now become infected. Also,  Eq.\,\ref{removed} takes into account the ratio between infected people and the duration of the infection, also called incubation time, being such a ratio proportional to the number of people recovering from the disease. This factor, namely $I'/D$, is negative in Eq.\,\ref{infected}, as expected, since recovered people are no longer part of the group of infected people. As previously mentioned, the set of three equations enables us to obtain the epidemic curve for an infectious disease.
\begin{figure}[h!]
\centering
\includegraphics[width=0.98\textwidth]{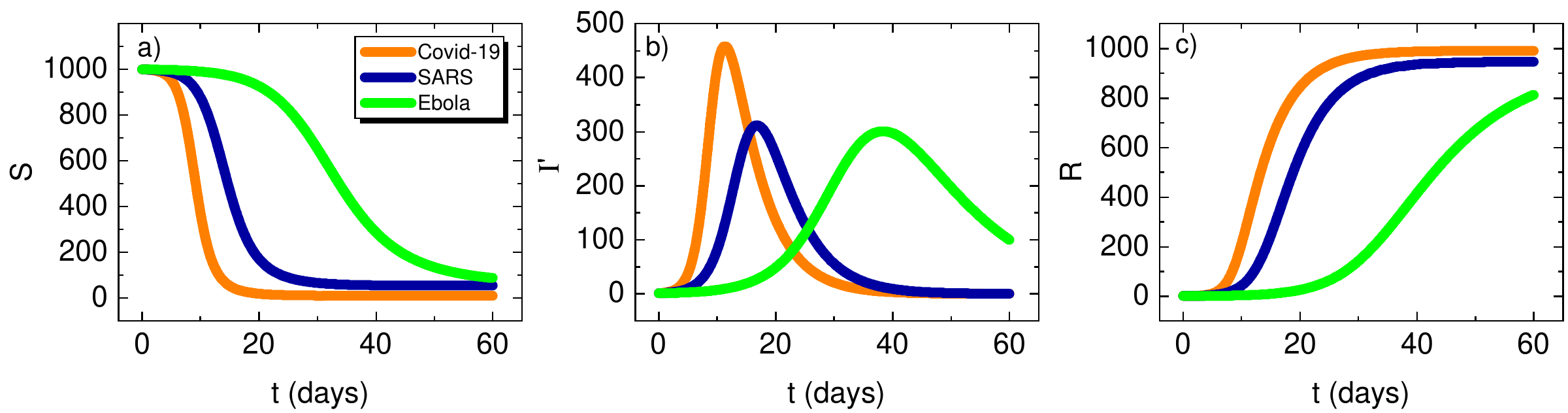}
\caption{\footnotesize Epidemic curves for Covid-19 (orange color solid line), SARS (navy solid line), and Ebola (green solid line). a) Number of susceptible people $S$, b) number of infected people $I'$, and c) number of immune people $R$ \emph{versus} time $t$, employing the parameters $R_0$ (number of people directly infected) and $D$ (incubation time) for each disease here considered taken from Refs.\,\cite{R0covid19,R0SARS,R0ebola}. Details in the main text.}
\label{Fig-1}
\end{figure}
In order to discuss the SIR model and its respective generated epidemic curves in a comprehensive way, we have applied such a model for Covid-19, SARS, and Ebola in order to discuss the implications of the disease spread for distinct $R_0$ factors (Fig.\,\ref{Fig-1}). Note that the higher value of $R_0$ is associated with Covid-19, most likely due to its relatively easy infection because of the furin enzyme. Figure\,\ref{Fig-1} shows the epidemic curves employing the parameters $R_0$ and $D$ reported in the literature for the case of Covid-19 \cite{R0covid19}, SARS \cite{R0SARS}, and Ebola \cite{R0ebola}. The number of susceptible people $S$ in panel a) of Fig.\,\ref{Fig-1} for Covid-19 decreases more rapidly than SARS and Ebola, which is a direct consequence from the fact that Covid-19 presents a higher $R_0$ factor when compared with the other diseases. As a consequence, the number of infected people over time depicted in panel b) of Fig.\,\ref{Fig-1} increases more rapidly and thus the number of immune people $R$ from Covid-19 in panel c) of Fig.\,\ref{Fig-1} increases more slowly due to the quick spread of the disease. In summary, Fig.\,\ref{Fig-1} shows the typical epidemic curves for each disease spread here considered, in the frame of the SIR model.
Note that the SIR model, through the knowledge of $R_0(\beta', \kappa, D)$, enables us to infer predictions about the spread of the disease.
According with recent works, environmental factors such as the weather, can also be taken into account when treating a disease spread \cite{Xiao2020}. Such an analysis is key in predicting the epidemiologic curves during seasonal changes in some countries. Essentially, environmental factors, i.e.,  temperature changes, can impact on the floating time of respiratory droplets in the atmosphere, leading thus to an enhancement of the infection probability. Such an influence of the environment on the spread of the disease can be  associated with the $\beta'$ factor in the light of the SIR model.
In practical terms, several factors enable us to minimize the risk of infection. For example, in the context of the SIR model, upon joining the quarantine both $\beta'$ and $\kappa$ are reduced, since the number of contacts between people is lowered in the same way as the probability of being infected. On the other hand, upon going to the supermarket using a face mask, for instance, $\beta'$ can be lowered but not necessarily $\kappa$ since there will still be contact between people. Such actions are not only important on containing the disease spread, but also are key in preventing the resurgence, or small new outbreaks.
In a broader context, other mathematical models can be used to investigate the spread of diseases, such as MSEIR, MSEIRS, SEIR, SEIRS, SIRS, SEI, SEIS, SI, and SIS, where M is passively immune infants and E is the exposed people in the latent period \cite{Hethcote2000, Grassly}. However, as pointed out in Ref.\,\cite{Julia2020}, the simple application of the SIR model, for instance, is naive and does not suffice to describe the spread of Covid-19. Given the various involved factors, a broader analysis considering, for instance, social aspects and urbanization, is required. Such aspects are discussed in Section \ref{WN}.

\subsection{The Ising-model and the Covid-19 outbreak}
It is evident that the mathematical models aforementioned used to describe the spread of epidemics are not built based on concepts of Statistical Physics. However, an adaptation of the Ising-model can be made to describe the Covid-19 spread. Hence, in what follows we discuss the celebrated Ising-model \cite{Ising}. Although initially proposed for describing magnetic systems, over the last decades the Ising-model has been revealed as an appropriate tool to describe several phenomena. Indeed, this includes the supercooled phase of water \cite{SR}, the vicinity of the Mott critical end point \cite{Barto,2015, PRL2020}, magnetic field-induced quantum critical points \cite{PRB2019},  econophysics  \cite{econophysics}, democratic elections \cite{Bar-Yam2020,Jordan2020}, as well as the spread of diseases \cite{Pinol2012,McKendrick1927,Amaku2019,Cushing2008}, just to mention a few examples.
Before discussing the adaptation of the Ising-model for the description of the Covid-19 spread, we recall here the classical one-dimensional Ising-model \cite{Baxter,Nolting}. In order to make a brief review of such a model, we focus here on the case of applied longitudinal magnetic field. Essentially, the quintessence of the Ising-model lies on the assumption that magnetic moments are coupled only with their nearest neighbors, being the Hamiltonian of a linear chain of $N$ spins given by \cite{Nolting}:
\begin{equation}
H = - J_{i,i+1} \sum_{i=1}^{N} S_iS_{i+1} - B\sum_{i=1}^{N}S_i,
\label{ising}
\end{equation}
where $J_{i,i+1}$ is the coupling constant between neighboring magnetic moments on sites $i$ and $i+1$, $S_i$ ($S_{i+1}$) represents the spin on a site $i$ ($i+1$), and $B$ refers to the longitudinal applied magnetic field. At this point, it is worth emphasizing that the coupling constant  $J_{i,i+1}$ , also called magnetic exchange coupling constant, will be used to emulate the contact between infected and non-infected people in our approach.
Next, we make use of the  Ising-model \cite{Ising}, logistic function \cite{Brauer2012} and percolation theory (to be introduced in Section \ref{Perco}) to describe the spread of Covid-19. We demonstrate the close relation between the number of people following the proposed quarantine by the WHO (World Health Organization) and the spread of Covid-19.
Real examples of the effectiveness of the social distance can be found, for instance, in Brazil upon comparing the Covid-19 spread between Belo Horizonte - MG and S\~ao Paulo - SP. Since its outbreak in Brazil, the number of both new cases and fatalities reported for S\~ao Paulo city are still increasing up to date and represent the higher number of cases in the country \cite{nyt,bbc}. On the other hand, Belo Horizonte is still reporting on very low number of both new cases and fatalities due to Covid-19, demonstrating clearly the result of an early and very strick social distancing. Although it is quite obvious that the more people respect the quarantine the lower will be the number of infected people, a proper quantitative description is still lacking. Unfortunately, for many governmental leaders around the world, it is not obvious that the quarantine is crucial in containing the Covid-19 disease spread.
It is worth mentioning a recent analysis about the so-called \emph{Digital Herd Immunity}, reported in Ref.\,\cite{Bulchandani2020}, where there is a tracking of Covid-19 infected people through a smartphone applicative. Interestingly enough, the critical fraction of people using such applicative for the containment of the Covid-19 spread is about 75\% of the population. The latter means that if the entire population, infected or not, are tracked by the applicative and respect the social distancing, the spread of the disease is contained. This paper is organized as follows: in Section \ref{Section2}, we present a discussion about the adaptation of the Ising-model to the Covid-19 spread and deduce, in a pedagogical way, the logistic function, applying it in the description of both the number of infections and fatalities due to Covid-19. In Section \ref{LFL}, we present an analogy between the electron interaction in condensed matter Physics and the interaction between infected and non-infected people in epidemics. A comparison between the logistic function and the SIR model is presented in Section \ref{LogSIR}. In Section \ref{Perco}, we present a comprehensive review of the percolation theory, the Cayley tree, and the Bethe lattice aiming to describe the behavior of the Covid-19 spread in the context of percolation theory. Conclusions and perspectives are also presented.

\section{Covid-19 Spread, Infections and Fatalities}\label{Section2}

\subsection{The genesis of the \emph{interaction} parameter}
Under the light of condensed matter Physics, we consider that the \emph{interaction} (contact) $\delta\varepsilon$ between infected and non-infected people can be associated with the magnetic exchange interaction between nearest-neighbour magnetic moments in the well-established Ising-model \cite{PRL2020,SR,2019,Gene}, cf. discussions in the previous Section. Hence, analogously to the case of the Ising-model for magnetism with $N$ spins $S = 1/2$, we assume that the number of infected people $p_i = +1/2$ is $N^+$, while the number of non-infected people $p_i=-1/2$ is labelled by $N^-$, being $(N^+ + N^-) = N$ where $N$ is the total number of considered habitants. In other words, infected and non-infected people can be identified considering a  $S = 1/2$ Ising-like variable
$p_i = +1/2, -1/2$, so that we write $N^+=\sum_{i=1}^N 2\left(p_i+\frac{1}{2}\right) p_i$, and $N^-=\sum_{i=1}^N2\left(p_i-\frac{1}{2}\right)p_i$. The latter enable us to select mathematically infected and non-infected people, respectively.
Yet, we consider that two people who are infected by Covid-19 have no effect on each other, so that in this case the interaction $\delta\varepsilon = 0$ and, otherwise, $\delta\varepsilon \neq 0$. In the same way, a non-infected person has no effect in another non-infected person. Essentially, in our approach $\delta\varepsilon$ quantifies, at some extent, the key role played by the quarantine in the spread of Covid-19. Note that $\delta\varepsilon$ plays a similar role than $\kappa$ in the frame of the SIR model, discussed previously.  Following a similar mathematical treatment reported by us elsewhere, cf.\,Ref.\,\cite{PRL2020}, we write a mathematical function to describe the total population $C_{T}$ taking into account the contact between non-infected and infected people, which is emulated by $\delta\varepsilon$:
\begin{equation}
C_{T}=C_h -4\delta\varepsilon\sum_{i\neq j =1}^N \left[\left(\frac{1}{2}+p_i \right)\left(\frac{1}{2}-p_j \right) + \left(\frac{1}{2}-p_i \right)\left(\frac{1}{2}+p_j \right) \right] p_i p_j,
 \label{contamination}
\end{equation}
where $C_h$ is the total number of healthy people before the spread of the virus and $p_j$ refers to a neighbour person of $p_i$. Note that the second term of the right side of Eq.\,\ref{contamination} will always be negative for $p_i = +1/2$ and $p_j = -1/2$ and vice-versa. Although the various applications of the Ising-model to distinct research areas are known in the literature, as discussed in the Introduction section, Equation\,\ref{contamination} provides a new way of mathematically selecting infected ($p_i = +1/2$) and a non-infected person ($p_j = -1/2$) taking into account the \emph{interaction} parameter $\delta\varepsilon$. Furthermore, Eq.\,\ref{contamination} indicates that, when $\delta\varepsilon \neq 0$, i.e., infected people interact with healthy people, $C_h$ is decreased.
Evidently, the total number of the population remains constant. It is worth mentioning that even in a hypothetical situation in which all the population joins the quarantine, there would still be a small finite $\delta\varepsilon$ originated from household interactions, which are usually longer and more frequent and thus, under certain circumstances, can lead to new infections. In order to determine a proper mathematical expression for $\delta\varepsilon$, one must take into account the number of isolated people $n$ in quarantine. We propose that the contact between infected and non-infected people can be described considering
$\delta\varepsilon \propto n^{-1/2}$, with 1 $<  n <  N$. As a matter of fact, we have employed different power-laws for $\delta\varepsilon$ in terms of $n$, but it turns out that  $\delta\varepsilon \propto n^{-1/2}$ provides an appropriate fit for the data set. This is a reasonable assumption, since as the number of people in quarantine decreases, the contact between them increases, which, in the present scenario, is represented mathematically by an increase in $\delta\varepsilon$. Note that Eq.\,\ref{contamination} does not incorporate any time evolution.

\subsection{The Gaussian-type function and Gaussian processes}
It turns out that upon analyzing already officially reported epidemic curves \cite{Hollingsworth2020,Earn2014} for Covid-19, their typical time evolution in terms of new cases \emph{per} day follows an initial rapid increase of the number of infected people, which is usually assumed to be exponential.  After such initial exponential growth, the number of infected people reaches a threshold value and starts decreasing monotonically as time continues to evolve. Essentially, in qualitative terms, the typical shape of epidemic curves is more or less the one of a simple Gaussian-type function. At this point, it is worth mentioning that although the epidemics data set here discussed follows the behavior of a Gaussian function, our analysis has nothing to do with a probability distribution. For the sake of completeness, we recall the mathematical expression for the Gaussian function:
\begin{equation}
y(t) = y_0 + \frac{A}{w\sqrt{\frac{\pi}{2}}}e^{-2\frac{(t-t_c)^2}{w^2}},
\end{equation}
where $y_0$ is related with an initial value, $A$ is a normalization constant associated with the area under the curve, $t$ is the time, $t_c$ is the value of $t$ associated with the maximum value of $y(t)$, and $w = 2\sigma$, where $\sigma$ is the standard deviation.
Hence, following a similar analysis employed by the authors of Ref.\,\cite{Bar-Yam2020}, we make use here of a simple Gaussian function to describe the spread of Covid-19. In our analysis, the interaction between infected and non-infected people is incorporated in the Gaussian function by summing up $n^{1/2} \propto \delta\varepsilon^{-1}$ in $t_c$ and $w$ into the Gaussian function. The incorporation of $\delta\varepsilon$ in terms of the probability of a person being infected is discussed into details in Subsection \ref{Ising-Bethe}. We stress that in our analysis, we have summed up $\delta\varepsilon^{-1}$, as previously described, based on the argument that when the number of people $n$ in quarantine is increased, the contact between infected and non-infected people is reduced and, as a consequence, $\delta\varepsilon^{-1}$ is increased. Essentially, we have summed up $\delta\varepsilon^{-1}$ as follows:
\begin{equation}
y(t) = y_0 + \frac{A}{(w+\delta\varepsilon^{-1})\sqrt{\frac{\pi}{2}}}e^{-2\frac{[t-(t_c+\delta\varepsilon^{-1})]^2}{(w+\delta\varepsilon^{-1})^2}}.
\end{equation}
This gives rise to a broadening of the Gaussian distribution function, so that $w$ is increased  and its maximum is reduced. Note that by doing so, the area under the curve remains constant. It is worth emphasizing that our analysis based on the incorporation of $\delta\varepsilon$ in the Gaussian function was only possible due to the Ising-like model that we have introduced previously, within which $\delta\varepsilon$ has its genesis. We emphasize that the modeling of the epidemic curves is performed by assuming that there are several additional parameters ruling the global behavior of the time evolution of the key factors incorporated in the SIR model, i.e., the number of infected, recovered and susceptible people. However, it is important to reinforce that a real disease spread depends on an ensemble of parameters, which are not usually taken into account to describe the global dynamics of the epidemic curves. These additional factors  cannot be modelled by employing solely elementary analytic functions. Thus, in such cases, Gaussian processes may be implemented in the model to increase the accuracy of best-fit solutions describing the time evolution of the epidemics parameters.
Essentially, Gaussian processes can be understood in the following way: suppose we have a given phenomenon (in our case, the number of infected ($I'$), recovered ($R$) and susceptible ($S$) people in the light of an epidemic) to be modeled as a function of time $t$. There is a set of well-determined parameters which are responsible for dictating the predominant behavior regarding the time evolution of these particular parameters, cf.\,Eqs.\,\ref{susceptible}, \ref{infected}, and \ref{removed}. However, other factors may play a minor, although non-negligible, role in modeling the epidemic curve. A simple example of a factor which could change the epidemic curve is the weather. It is known that weather prediction is one of the most difficult tasks in modern predictive analysis due to the intrinsically stochastic aspect of this phenomenon. If the virus corresponding to a given epidemic presents a ``weakness'' to, e.g., hot weather, an abrupt change in the temperature may cause small fluctuations in the number of infected people over a short period of time. Such a period of time is comparable to the timespan  in which the temperature remains relatively high. Supposing that these temperature fluctuations take place for some days in a given timespan, one has a stochastic variable adding a ``noise'' to the behavior, for instance, of the number of susceptible people over time $S(t)$. The Gaussian processes is thus an algorithm for introducing these kind of random fluctuations in the epidemic model. The so-called Gaussian aspect is attributed to the way in which the noise is added. It is known that a sequence of randomly generated points can be approximated by a Gaussian distribution function for large values of the number of data points, being such result known as the Central Limit Theorem. Thus, the incorporation of randomly generated fluctuations in the deterministic portion of $S(t)$ implies a more complete modeling. Indeed, this is true not only for the deterministic nature of $S(t$), already described by the SIR model, but also to the noise introduced by, for instance, the weather variation and its impact on the survival of the virus causing the given disease. The goal is, eventually, to determine which value of standard deviation $\sigma$ better suits the Gaussian processes, i.e., the one which minimizes the deviation between the actual data of the epidemic curve and the model curve described by $S(t) = S(t)_{det} + S(t)_{gauss}$, where $S(t)_{det}$ is the deterministic part of the function $S(t)$ and $S(t)_{gauss}$ is the stochastic contribution to the function $S(t)$. Of course, such modelling requires a more complex implementation.
In Ref.\,\cite{Mackey} an example of a public available code to model a Gaussian process in one dimension is reported. Note, however, that such procedures are random in nature and serve only the purpose of increasing the accuracy of the fits by taking into account a detrending process. We thus mention here the existence of such methods, but we do not implement them in the current work, since no additional insights can be gained in the discussions related to the spread of Covid-19.

\subsection{The logistic and the Fermi-Dirac-like function applied to epidemics}
Given the relatively large amount of  available data for China and the achievement of a proper control of the spread of Covid-19 in its territory, in the following we pedagogically deduce the logistic function by making use of the also celebrated standard population growth model \cite{Stewart, Ricieri}. We assume a maximum number of possible infected people labelled by $P$; considering $I(t)$ the number of infected people at a given time $t$ and, as consequence, [$P - I(t)$] is the number of people that can be infected. Note that $[\frac{P - I(t)}{P}]$  corresponds to the percentage of the population that will be infected. Also, we consider that the number of people that can be infected by $I(t)$  in a time interval $\Delta t$ is $\left[\frac{P-I(t)}{P}\right]n^*\Delta t$, where $n^*$ quantifies how many times an infected person interacts with a non-infected person. The number of infected people in a time $(t + \Delta t)$ minus the number of infected people at $t$ is proportional to the number of people of the population that can be infected by the already existent infected people at a given time $t$. Hence, it is straightforward to write:
\begin{equation}
\frac{I(t+\Delta t) - I(t)}{\Delta t} = K\left\{\left[\frac{P - I(t)}{P}\right]n^* I(t)\right\},
\label{diseasespread}
\end{equation}
where $K$ is a non-universal proportionality constant. Upon taking the limit for $\Delta t \rightarrow 0$ in both sides of Eq.\,\ref{diseasespread}, we achieve the following differential equation:
\begin{equation}
\frac{dI}{dt} = C\left(\frac{P-I}{P}\right)I,
\label{diseasespread2}
\end{equation}
where $C = Kn^*$ quantifies the frequency of infection. Making the integration in both sides of Eq.\,\ref{diseasespread2}, we have:
\begin{equation}
I(t) = \frac{P}{me^{-Ct}+1},
\label{fit}
\end{equation}
where $m$ is a non-universal integration constant. Equation\,\ref{fit} has some reminiscence of the well-known Fermi-Dirac (FD) distribution function \cite{Kittel}. Interestingly, Eq.\,\ref{fit} is also typically employed in chemical kinetics \cite{kinetics}, anaerobic biodegradability tests \cite{anaerobic}, and even on germination curves \cite{germination}. It provides a robust description for the evolution of the number of infected people $I(t)$ over time, saturating at $P$. Also, Eq.\,\ref{diseasespread2} can be recognized as the well-established logistic differential equation \cite{Stewart}, which has a general solution of the type \cite{Kyurkchiev}:
\begin{equation}
g(x) = \frac{L}{e^{-k(x-x_0)}+1},
\label{gx}
\end{equation}
where $k$ is the growth rate of the function, $L$ is the maximum number of $g(x)$, and $x_0$ is the midpoint value of the function. Thus, Eq.\,\ref{fit} can be rewritten in a similar way of Eq.\,\ref{gx}:
\begin{equation}
I(t) = \frac{P}{e^{-C(t-t_0)}+1}.
\label{mt0}
\end{equation}
Note that Eq.\,\ref{mt0} is equivalent to Eq.\,\ref{fit} if we consider $m = e^{Ct_0}$. The logistic function can also be written in the trigonometric form as the distribution function:
\begin{equation}
\frac{I(t)}{P} = \frac{1}{2}+\frac{1}{2}\tanh{\left[\frac{C(t-t_0)}{2}\right]}.
\label{distributionfunction}
\end{equation}
Although Eq.\,\ref{fit} is essentially the well-known logistic function, we will consider it under the view of condensed matter Physics as a Fermi-Dirac-type (FD-type) function. This is because of their mathematical similarities.  The data set for  Covid-19 here discussed are available in Ref.\cite{worldometers}. It is worth mentioning that we have focused our analysis in countries presenting a more advanced picture of the Covid-19 spread, such as South Korea and China \cite{worldometers}.

\subsection{Data analysis and discussion for the case of infections}
\begin{figure}[h!]
\centering
\includegraphics[width=0.5\textwidth]{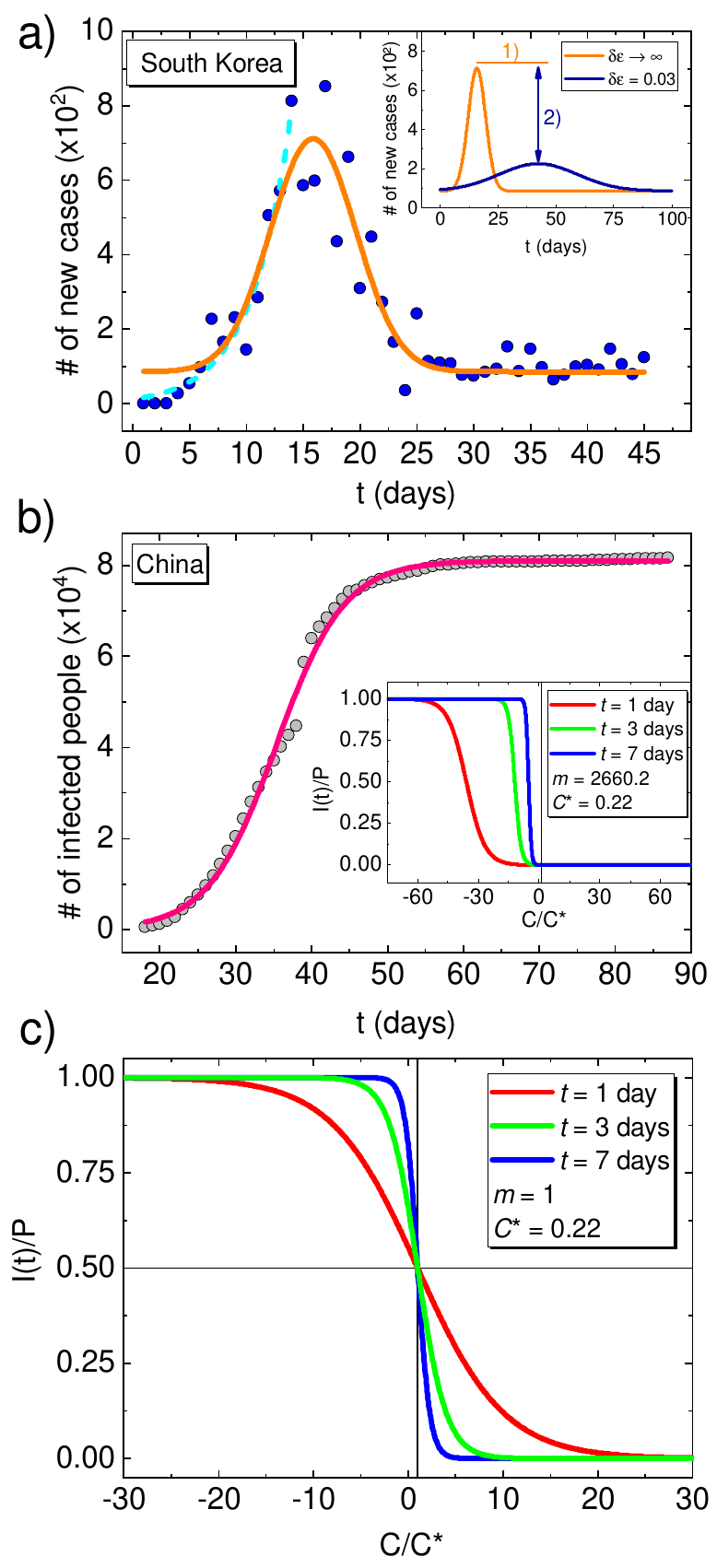}
\caption{\footnotesize \textbf{a)} \textbf{Main panel}: number of new cases \emph{versus} time (in days) for Covid-19 in South Korea, fitted with a Gaussian (orange color solid line) and exponential (dashed cyan line) functions. \textbf{Inset}: corresponding Gaussian fit for $\delta\varepsilon \rightarrow \infty$ (orange color solid line) and $\delta\varepsilon = 0.03$ (navy blue solid line);  1) delimitates  the maximum number of new cases for $\delta\varepsilon \rightarrow \infty$ and 2) the reduction of the number of new cases when the number of people in quarantine is increased \cite{Media}. \textbf{b)} \textbf{Main panel}: accumulative number of infected people \emph{versus} time (in days) of Covid-19 for China and the correspondent fitting (pink solid line) employing Eq.\,\ref{fit}. \textbf{Inset}: $I(t)/P$ \emph{versus} $C/C^*$ with $m = 2660.2$ and $C^*$ = 0.22 (obtained from the fitting of the data set shown in the main panel)
for 1 day (red solid line), 3 days (green solid line), and 7 days (blue solid line).  \textbf{c)} $I(t)/P$ \emph{versus} $C/C^*$ with $m = 1$ and $C^*$ = 0.22 for 1 day (red solid line), 3 days (green solid line), and 7 days (blue solid line). % employing Eq.\,\ref{fit}.
For the sake of comparison, we have employed $C^*$ obtained in the fitting of the data for China in b).
Data set available in Ref.\,\cite{worldometers}.}
\label{Fig-3}
\end{figure}

We start our analysis recalling the data set for Ebola, SARS, and Influenza A/H1N1.
\begin{figure}[h!]
\centering
\includegraphics[width=0.5\textwidth]{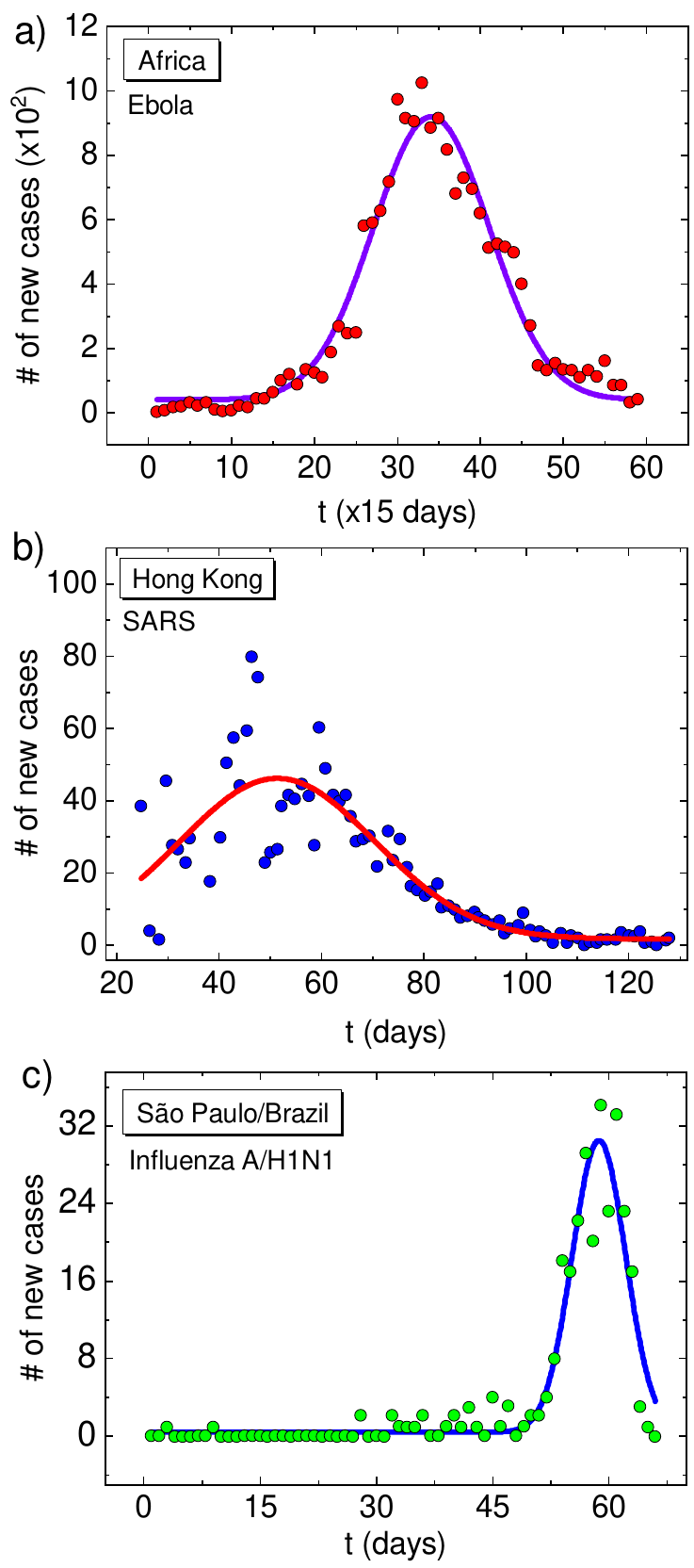}
\caption{\footnotesize  Number of new cases \emph{versus} time in days for: \textbf{a)} Ebola \cite{ebola},\textbf{ b)} SARS \cite{SARS}, and \textbf{c)} Influenza A/H1N1 \cite{H1N1}. The solid lines in all panels represent the Gaussian fitting of the data set for each case.}
\label{Fig-2}
\end{figure}
Their epidemic curves, depicted in Fig.\,\ref{Fig-2}, can be described by a fitting employing a Gaussian function.
\begin{table}[!h]
\footnotesize
\centering
\begin{tabular}{|c|c|c|c|c|c|}
\hline\hline
\textbf{Disease}                & $y_0$ & $A$       & $w$     & $t_c$ & $\sigma$ \\ \hline\hline
\textbf{Ebola}                  & 41    & 15418  & 13.98 & 34.08 & 6.99    \\ \hline
\textbf{SARS }                  & 2     & 2122   & 38.08 & 51.39 & 19.04    \\ \hline
\textbf{Influenza A (H1N1) }    & 0.42  & 260  & 6.89  & 58.69 & 3.44     \\ \hline
\textbf{Covid-19 (South Korea}) & 85.94    & 5890.41    &  7.51  & 15.88 & 3.75    \\ \hline
\textbf{Covid-19 (Austria)}     & 12.82 & 4317.33 & 8.34  & 24.24 & 4.17    \\ \hline
\end{tabular}
\caption{\footnotesize Gaussian fitting parameters for several epidemics [see Figs.\,\ref{Fig-2} and \ref{Fig-3} a)]. The existence of other external parameters influencing the time evolution of the number of new cases causes fluctuations on the data. Since such factors can be described, in general, by a Gaussian process, the associated errors of the fitting with respect to the real data are incorporated in the standard deviations of the fitting presented in Figs.\ref{Fig-2} and \ref{Fig-3}. Details in the main text.}
\label{tableparameters}
\end{table}
\normalsize
Table\,\ref{tableparameters} shows the parameters obtained in the fitting for the various epidemics (Fig.\,\ref{Fig-2}) and their respective standard deviations. At this point, it is worth emphasizing that it is not possible to make a forecast of an epidemic curve by only employing a Gaussian fitting of the data set associated with the initial growth of the epidemic curve. Now, we focus on the analysis of the spread of Covid-19. We start with the available data set for Covid-19 in South Korea, see Fig.\,\ref{Fig-3} a).
The number of people joining the quarantine is directly associated with the previously defined \emph{interaction} (contact) $\delta\varepsilon$ between infected and non-infected people. As discussed, in a hypothetical situation where no one joins the quarantine $n \rightarrow 0$ and so $\delta\varepsilon$ achieves its maximum value, which means that everyone interacts freely with each other. This would be the worst case at all.
It has been broadly discussed that upon increasing the number of infected people in quarantine, the maximum in the epidemic curve associated with the number of new cases is not only lowered, but it is also shifted, indicating that the spread of Covid-19 will be more contained. The latter corresponds to the desired situation in terms of controlling the spread, since the health government agencies will have more time to manage the situation. This is one of the reasons why a proper registration of the epidemic data is crucial. Note that our analysis based on the Ising-like model incorporates the key role played by social distancing, broadly discussed in the media, see, e.g., Ref.\,\cite{Media}. We demonstrate such a situation employing the data set available for South Korea, assuming a hypothetical finite $\delta\varepsilon$ incorporated in the Gaussian function employed in the analysis, cf. depicted in the inset of Fig.\,\ref{Fig-3} a). More specifically, when $\delta\varepsilon$ is lowered the maximum number of infected people is decreased and its position in time is shifted, making thus the disease spread more controllable. Such a decrease in $\delta\varepsilon$ represents more people joining the quarantine and thus avoiding contact with each other.  Interestingly, the number of infected people $I'$ over time in the frame of the SIR model, depicted in Fig.\,\ref{Fig-1} b), for the various diseases resembles the behavior of the number of new cases over time shown in the inset of Fig.\,\ref{Fig-3} for Covid-19. Such a resemblance of the role played by $\delta \varepsilon$ and $R_0$ is due to the fact $R_0$ depends on $\kappa$, which in turn represents an average interaction between people in a similar way than $\delta \varepsilon$. Indeed, the $R_0$ factor can be written in terms of $\kappa$, namely  $R_0 = \beta'\kappa D$ \cite{Giesecke}. Since $R_0$ is connected with the average number of contacts between people as a function of time $\kappa$, upon decreasing $R_0$ as shown in Fig.\,\ref{Fig-1} b) the shape of the infected people over time is broadened in the same way as in Fig.\,\ref{Fig-3} a) when $\delta\varepsilon$ is lowered. Essentially,  both behaviors strengthen the importance of respecting the social distancing and thus minimizing the average number of contacts between people. Now, we treat the available data set of the spread of Covid-19 in China in terms of the FD-like distribution function.  As depicted in Fig.\,\ref{Fig-3} b), the number of accumulated infected people \emph{versus} elapsed time starts to saturate after roughly 50 days after the outbreak of Covid-19. In other words, for large values of $t$ the ratio $I(t)/P \rightarrow 1$. Equation \ref{fit} fits nicely such data set, cf.\,Fig.\,\ref{Fig-3} b). Note that the number of infected people over time for China is marked by a small outbreak at $t \approx  40$ days, cf. Fig.\,\ref{Fig-3} b). Such a small outbreak can be due to several distinct factors, such as the flexibilization of social distancing or even due to a lack of attention of the population in taking the basic hygiene cares.
In the following, we discuss into more details the similarity between Eq.\,\ref{fit} and the FD distribution function. Rearranging Eq.\,\ref{fit}, we have:
\begin{equation}
\frac{I(t)}{P} = \frac{1}{me^{(C-C^*)t}+1},
\label{FD}
\end{equation}
being $C^*$ a constant introduced to play the role of the Fermi energy for the electron gas \cite{Kittel}. Note that $I(t)/P \leq 1$ and, for $m$ = 1 we have exactly the same form of the FD distribution function $f(E,T)$ \cite{Kittel}, where $E$ and $T$ refer, respectively, to the energy and temperature. In the present case, $C$ plays a role analogous to the energy for the Fermi gas, while $t$ is analogous to Boltzmann factor $\beta = 1/k_BT$, where $k_B$ is Boltzmann constant. Indeed, the time evolution of the Covid-19 spread has a similar significance than $T$ for the FD distribution function for the Fermi gas. The behavior of $I(t)/P$ as a function of $C/C^*$ is shown in the inset of Fig.\,\ref{Fig-3} b) for several values of $t$. Considering that in our fit using Eq.\,\ref{fit} for the spread of Covid-19 in China [inset of Fig.\,\ref{Fig-3} b)] we have obtained $m = 2660.2$, it becomes clear that we are dealing with a distorted version of the FD function where $m = 1$, cf.\,Fig.\,\ref{Fig-3} c). Hence, at some extent, we are faced with a behavior analogous to the distorted FD distribution for electrons in the picture of Landau Fermi-liquid (FL) \cite{fermiliquid,Pines}. The latter will be discussed into more details in Section \ref{LFL}. Also, we emphasize that the ratio $I(t)/P$ gives us the probability of a person being infected in a certain time $t$ in a similar way that $f(E,T)$ dictates the probability of a state with energy $E$ being occupied at a certain temperature $T$. We anticipate that the spread of Covid-19 for other countries, not discussed in the present work given the lack of available data, should follow the same behavior as here discussed for South Korea and China. The position of the maximum in the epidemic curve, described by a Gaussian function, as well as $I(t)$, will be a direct reflex of the policies taken by the health government agencies by a particular country.

\subsection{The logistic and the Fermi-Dirac-like function for the case of fatalities}
As reported by the Centers for Disease Control and Prevention (CDC) \cite{CDC}, there are some risk factors that increase the chance of an infected people to pass away due to infection by Covid-19. Such factors include mainly being with age above 65\,years old and having a serious underlying medical condition. In our approach, we refer to this set of factors as a global (Risk) $R'$-factor. We assume that the number of infected people presenting the $R'$-factor is $C_{R'}$ and the number of infected people without presenting the $R'$-factor $C_W$.  It is evident that the total number of infected people $C_I$ is given by $C_I = (C_W + C_{R'}$). Also, it is expected that people presenting the $R'$-factor are more likely to pass away and thus to reduce the total number of infected people. As discussed in the following, the number of new fatalities \emph{per} day follows a Gaussian fitting in the same way as the number of new infection cases \emph{per} day. Now, we focus on the number of accumulative fatalities over time. Based on the fact that the number of accumulative fatalities over time follows a behavior similar to the number of accumulative infected people over time, we carry out a similar analysis applied to the number of infected people over time, cf.\,previously discussed. The number of people that can pass away in a time interval $\Delta t$ is given by $\left[\frac{F - D(t)}{F}\right]d^{*}\Delta t$, where $F$ is the maximum number of fatalities, $D(t)$ the number of people that pass away in a certain time $t$, and $d^{*}$ is a parameter associated with the fatalities rate. Thus, the number of fatalities in a time interval $(t + \Delta t)$ minus the number of fatalities in a time $t$ is proportional to the number of infected people presenting the $R'$-factor. These assumptions enable us to write:
\begin{equation}
\frac{[D(t+\Delta t) - D(t)]}{\Delta t} = K'\left\{\left[\frac{F - D(t)}{F}\right]d^{*}D(t)\right\},
\end{equation}
where $K'$ is a proportionality constant. Thus, following the same elementary mathematical treatment as in the last Subsection, we can write:
\begin{equation}
D(t) = \frac{F}{m'e^{-C't}+1},
\label{numberofdeaths}
\end{equation}
where $m'$ is a non-universal integration constant and $C' = K'd^{*}$ is the fatality rate. Equation\,\ref{numberofdeaths} provides a reasonable description of the number of accumulative fatalities over time and resembles, as stated previously, a distorted Fermi-Dirac-like distribution function (logistic function), being that for the ideal case $m' = 1$. Interestingly, the number of accumulative fatalities over time, described by Eq.\,\ref{numberofdeaths}, also follows a logistic function in the same way as the number of accumulative infections over time, cf. Eq.\,\ref{fit}. Next, we discuss the fitting of the number of accumulative fatalities in China employing Eq.\,\ref{numberofdeaths}.

\subsection{Data analysis and discussion of the fatalities and the infection capacity}
\begin{figure}[h!]
\centering
\includegraphics[clip,width=0.5\columnwidth]{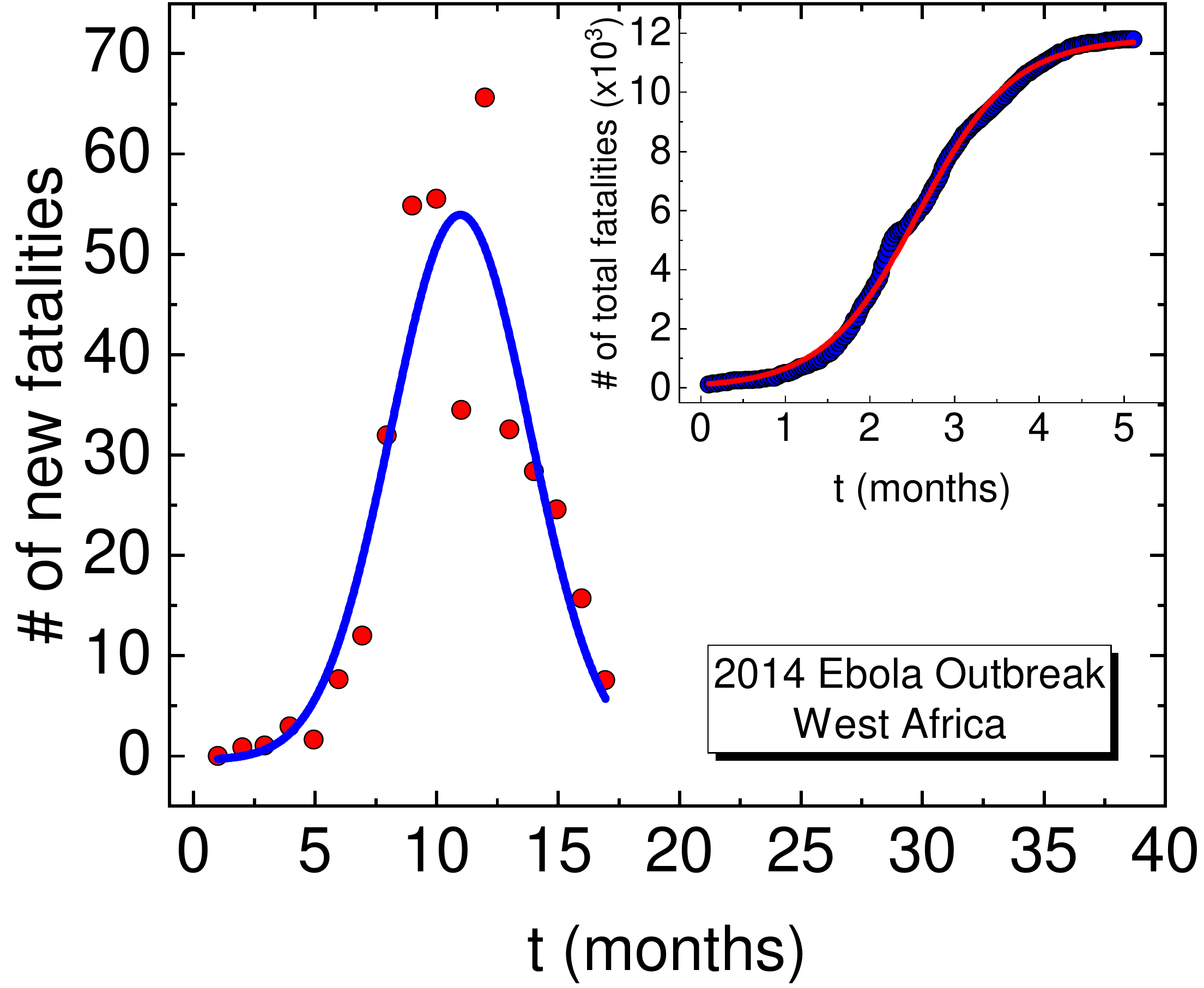}
\caption{\footnotesize Main panel: number of new fatalities \emph{per} month in the 2014 Ebola outbreak in West Africa fitted with a Gaussian function (blue solid line),  being the obtained fitting parameters displayed in Table \ref{tableparameters_SM}. Inset: total number of fatalities \emph{per} month during the Ebola outbreak fitted (red solid line) with Eq.\,\ref{numberofdeaths}, being $F =$ 11817.12, $m' =$ 100.61 and $C^* =$ 1.80. Data set available in Ref.\,\cite{WHO}.}
\label{Fig-4}
\end{figure}

\begin{figure}[h!]
\centering
\includegraphics[clip,width=0.5\columnwidth]{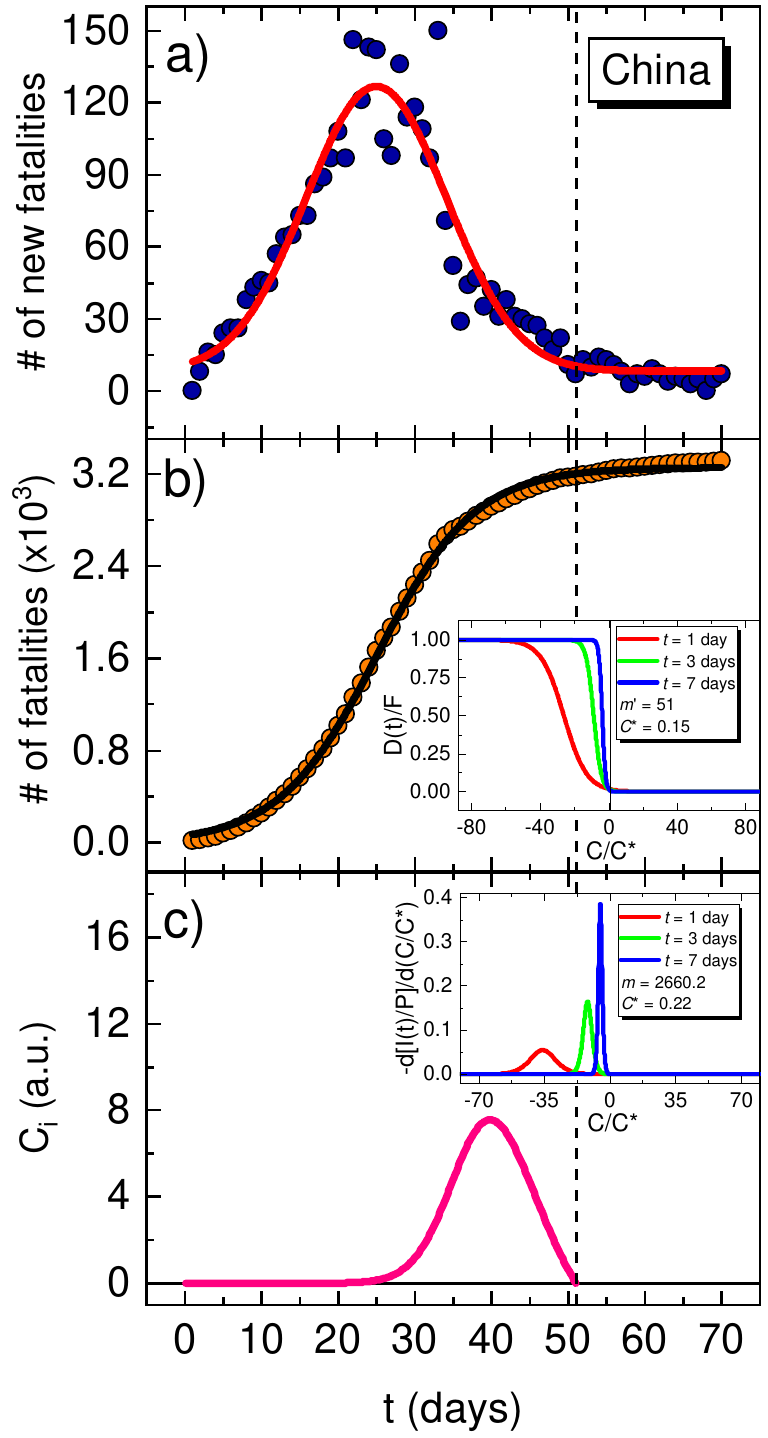}
\caption{\footnotesize \textbf{a)} Number of new fatalities \emph{per} days of Covid-19 in China fitted with a Gaussian function (red solid line). \textbf{b)} Main panel: accumulative number of fatalities \emph{per} days in China fitted (black solid line) with Eq.\,\ref{numberofdeaths}. Inset: $D(t)/F$ as a function of $C/C^*$ showing a distorted Fermi-Dirac-like behavior employing $m' = 51$ and $C^* = 0.15$.
\textbf{c)} Main panel: infection capacity $C_i$ (in arbitrary units) as a function of time $t$ in days (pink solid line) computed employing the fitting parameters for China, namely $C^*=0.22$, $m$ = 2660.2, and $P = 80994.72$, using Eq.\,\ref{infectioncapacity}. Inset: $-d[I(t)/P]/d(C/C^*)$ \emph{versus} $C/C^*$ employing the fitting parameters for China for 1 day (red solid line), 3 days (green solid line), and 7 days (blue solid line). The vertical dashed indicates the $t \sim$ 50 days. Data set available in Ref.\,\cite{worldometers}.}
\label{Fig-5}
\end{figure}

Before starting the discussion regarding the data set for China, aiming to demonstrate the applicability of our approach, we recall the behavior of both the number of new fatalities and the total number of fatalities over time for the Ebola outbreak in West Africa in 2014 \cite{WHO}. As can be seen in the main panel of Fig.\,\ref{Fig-4}, the time evolution of the number of new fatalities follows a Gaussian function, whereas the total number (accumulative) of fatalities (inset of Fig.\,\ref{Fig-4}) is well-described by Eq.\,\ref{numberofdeaths}. Such a behavior is also followed by other pandemics  (not shown) \cite{Ritchie2020, Arcila2019} and thus reinforcing that we are dealing with an universal description of the epidemic curves of any disease.
Considering the number of new fatalities by Covid-19 for the specific case of China [Fig.\,\ref{Fig-5} a)], a similar behavior than the 2014 Ebola outbreak in West Africa (main panel of Fig.\,\ref{Fig-4}) is observed. As depicted in Fig.\,\ref{Fig-5} a), the number of new fatalities over time follows a Gaussian function as well. The total number of fatalities over time [Fig.\,\ref{Fig-5} b)] follows a behavior which resembles the one of a distorted FD-type function, cf.\,inset of Fig.\,\ref{Fig-5} b). Interestingly, the value of the parameter $m' = 51$ is much lower than the one for the accumulative number of infections over time, which is suggestive that the total number of fatalities \emph{per} day follows a less distorted FD-like distribution function.
\begin{table}[!h]
\footnotesize
\centering
\begin{tabular}{|c|c|c|c|c|c|}
\hline\hline
\textbf{Disease}                & $y_0$ & $A$       & $w$     & $t_c$ & $\sigma$ \\ \hline\hline
\textbf{Ebola}                  & $-$0.41    & 389.66  & 5.72 & 10.97 & 2.86    \\ \hline
\textbf{Covid-19}                  & 8.28     & 2725.28   & 18.35 & 24.95 & 9.17    \\ \hline
\end{tabular}
\caption{\footnotesize Gaussian fitting parameters for the number of new fatalities \emph{per} day for Ebola and Covid-19 [see main panel of Fig.\,\ref{Fig-4} and Fig.\,\ref{Fig-5} a)].}
\label{tableparameters_SM}
\end{table}
Also, under the perspective of condensed matter Physics, given the similarity between Eq.\,\ref{FD} and the FD distribution function \cite{Kittel}, a direct analogy can also be made for the density of states (DOS), i.e., $\frac{dN'}{dE}$, where $N'$ is the number of particles (electrons) and $E$ the energy in the case of the Fermi gas. Here, the proposed equivalent quantity is $\frac{dI(t)}{dC}$. For the sake of completeness, we recall the expression used in the calculation of the specific heat $C_{FG}$ for the Fermi gas \cite{Kittel}:
\begin{equation}
C_{FG} = \int_0^{\infty} dE\frac{dN'}{dE}(E - E_F)\frac{df(E,T)}{dT},
\end{equation}
where $E_F$ is the Fermi energy.
At this point, it is worth mentioning that the DOS for the Fermi gas is temperature independent \cite{Kittel}. Below we write the here proposed \emph{infection capacity} $C_i$:
\begin{eqnarray}
C_i&=&-\int_0^{C^*} dC \frac{dI(t)}{dC} (C-C^*) \frac{d[I(t)/P]}{d(1/t)} + \nonumber\\
&& \label{infectioncapacity} - \int_0^{C^*}dC [I(t)/P] (C-C^*) \frac{d^2[I(t)]}{dC\,d(1/t)}.
\end{eqnarray}
Note that for $t \rightarrow$  0, based on previous discussions, $C$ should have its maximum value. Also note the presence of an additional term in Eq.\,\ref{infectioncapacity}.
Such a term emerges because in the present case the DOS is time dependent, being such a feature key in understanding the present results.  Equation \ref{infectioncapacity} quantifies the infection capacity by Covid-19 over time. In the main panel of Fig.\,\ref{Fig-5} c), we show $C_i$ \emph{versus} time for China employing the fitting parameters discussed for new infections, cf.\,Fig.\,\ref{Fig-5} a). A Gaussian function-like behavior is clearly observed. In the case of China, the infection capacity goes to zero at $t \sim 50$ days (dashed vertical line in Fig.\,\ref{Fig-5}), which is in agreement with both the vanishing of new fatalities \emph{per} day and also with the saturation of the accumulative number of people which passed away, cf. Figs.\,\ref{Fig-5} a) and b), respectively. Furthermore, the lowering of $C_i$, preceded by its vanishing, is also associated with the saturation of the accumulative number of infected people, which also takes place at $t \sim 50$ days. The inset of Fig.\,\ref{Fig-5} c) depicts $-d[I(t)/P]/d(C/C^*)$ \emph{versus} $C/C^*$. Interestingly, for $t \rightarrow \infty$ the number of people that can be infected is reduced and, consequently the contact (\emph{interaction}) between infected and non-infected people is reduced as well. Such a reduction leads to a behavior closer to an ordinary FD distribution function where no interactions are present, i.e., we approach a clean Dirac delta function.

\section{The Landau Fermi-Liquid picture}\label{LFL}
Following the discussions presented in the previous section, a similar situation is also found in a Landau FL electronic system \cite{Pines}. Upon bringing the quasi-particle (interacting) character to a non-interacting picture, the distorted FD function becomes a regular FD distribution.
In the frame of the Landau FL picture the equilibrium distribution function of the quasi-particles is given by \cite{Pines,Landau1987}:
\begin{equation}
n_\mathbf{p}^0 = \frac{1}{[e^{(\tilde{\epsilon}_\mathbf{p} - \mu)/k_B T} + 1]},
\end{equation}
where $\tilde{\epsilon}_\mathbf{p}$ is the quasi-particle local energy and $\mu$ the chemical potential. The quantity $\tilde{\epsilon}_\mathbf{p}$ depends on the interaction between the quasi-particles $f_{\mathbf{p},\mathbf{p'}}$ and can be written in the form $\tilde{\epsilon}_\mathbf{p} = \epsilon_\mathbf{p} + \sum_{\mathbf{p'}}f_{\mathbf{p},\mathbf{p'}}\,\delta n_{\mathbf{p'}}$, where $\epsilon_\mathbf{p}$ is the energy associated with the case when the interactions between quasi-particles are absent and $\delta n_{\mathbf{p'}}$ is the distribution function of the excited quasi-particles. Thus, ${n_\mathbf{p}}^0$ can be rewritten as follows:
\begin{equation}
n_\mathbf{p}^0 = \frac{1}{[e^{(\sum_{\mathbf{p'}}f_{\mathbf{p},\mathbf{p'}}\,\delta n_{\mathbf{p'}})/k_B T}e^{(\epsilon_\mathbf{p} - \mu)/k_B T} + 1]},
\end{equation}
where the term $e^{(\sum_{\mathbf{p'}}f_{\mathbf{p},\mathbf{p'}}\,\delta n_{\mathbf{p'}})/k_B T}$ can be recognized as the previously discussed factor $m$. In the case where $f_{\mathbf{p},\mathbf{p'}} = 0$ then $m = 1$, which leads to the case of a non-interacting Fermi-gas. On the other hand, when $f_{\mathbf{p},\mathbf{p'}} \neq 0$ we have a distorted Fermi-Dirac distribution function, which corresponds to the case of a Landau FL as previously discussed. At this point, an analogy can be made between the behavior of electrons in solids and the interaction between people during an epidemic. In fact, the interaction between quasi-particles $f_{\mathbf{p},\mathbf{p'}}$ plays a role analogous to the \emph{interaction} parameter $\delta\varepsilon$, which quantifies the contact between an infected and a non-infected person in the modelling of the Covid-19 spread.

\section{The logistic function \emph{versus} SIR model}\label{LogSIR}
In the following, we make a comparison between the logistic equation and the SIR model. Note that Eq.\,\ref{fit} represents the total number of infected people as a function of time $I(t)$. In the frame of the SIR model, knowing $R_0$ and $D$ we have access to both the number of new infections per day $I'(t)$ and accumulative total number of infections by calculating $\int I'(t) dt$. Hence, it is possible to obtain the accumulative total number of infected people in a similar way when employing the logistic equation. This demonstrates that, although it is appropriate for the description of epidemics, the SIR model presents a much more complex treatment than the application of the logistic equation, making it easier to determine the epidemic curve by the latter. It is clear that to make use of Eq.\,\ref{fit} one needs to know the infection rate and the constant $m$ ($\propto t_0$), while the SIR model considers purely epidemiological factors, such as $\beta'$ and $\kappa$. Thus, although simple, the logistical equation depends on factors that can only be determined after the end of the epidemic in a given place or region. Besides the use of the logistic function to describe sigmoidal growth curves, the so-called Gompertz function can also be employed \cite{Tabeau2001}. Essentially, the main difference between the logistic and the Gompertz function lies on the fact that for the Gompertz function the growth is more steep and it approaches the asymptote smoother than in the case of the logistic function. Thus, in the frame of the Covid-19 spread, the use of either the logistic or the Gompertz function will depend on the steepness of the total accumulative number of cases over time, which can be different for each country. Recently, a predictive model employing the Gompertz function, moving regression, and a so-called Hidden Markov Model was reported \cite{Utsunomiya2020}. In Ref.\,\cite{Utsunomiya2020}, the authors provide a numerical fitting of the epidemic data for all countries in order to predict the number of new cases in the next day, discussing also the disease spread in terms of four stages: the start of the outbreak (lagging), the initial exponential growth, the deceleration, and the stationary phase. Also, the authors of Ref.\,\cite{Utsunomiya2020} provide a real-time platform to monitor the growth acceleration of the Covid-19 spread in all countries aiming to measure the effectiveness of the mobility restrictions on the containment of such a disease.

\section{The Cayley tree and the Bethe lattice}\label{Perco}
As a natural continuation of previous sections, in what follows we present an analysis of the probability of a person being infected with Covid-19 in the frame of percolation theory. Over the past decades, the percolation theory \cite{Shlomo,percolation1,percolation2} has been used to describe the effects of disorder in superconductors \cite{superconductivity}, in the description of diluted magnetic semiconductors \cite{semiconductors}, in the analysis of traffic network \cite{Li}, and also in epidemiology \cite{epidemiology}, just to mention a few examples.
In condensed matter Physics, some powerful methods have been used to explore the electronic structure of systems of interest. Here, the density functional theory (DFT) and dynamical mean-field theory (DMFT) deserve to be mentioned. The latter enables the investigation of the so-called strongly correlated systems, i.e., systems in which interactions between electrons are taken into account and, as a consequence, the emergence of exotic phases. In a crude description, DMFT assumes an atom as an impurity in a matrix with several electrons under the influence of an effective field \cite{Kotliar2004}. Such an approximation becomes accurate as the coordination number ($z$) becomes infinite, i.e., it takes into account infinite nearest neighbors \cite{Metzner1989, Kotliar2004,Georges1994,Riordan2006}.
Within this context, the application of the Bethe lattice is appropriate.  Considering the application of percolation theory in the field of Solid State Physics, we highlight here the so-called Bruggemans's effective medium approximation \cite{Choy}.
\begin{equation}
x\frac{\varepsilon_a - \varepsilon_{eff}}{\varepsilon_{eff}+L(\varepsilon_a - \varepsilon_{eff})}+(1-x)\frac{\varepsilon_b - \varepsilon_{eff}}{\varepsilon_{eff}+L(\varepsilon_b - \varepsilon_{eff})}=0,
\end{equation}
where $x$  refers to the fraction of one of the phases of interest, $\varepsilon_a$ and $\varepsilon_b$ represent, respectively, the dielectric constant of the two distinct phases labelled $a$ and $b$, $\varepsilon_{eff}$ is the effective dielectric constant, and $L$ is the so-called shape factor.
In the frame of such an approximation, the goal is to analyse the dielectric responses of the two coexisting phases. Thus, when one of the phases achieves a critical concentration (or percolation threshold) one of the phases will dominate the Physics of the system significantly altering its dielectric response \cite{Efros}. Here, upon analyzing the spread of Covid-19 we associate the percolation probability with the probability of a person being infected in terms of the number of people in and out the quarantine. Figure \ref{Fig-7} shows the probability of infection as a function of  the number of connections $n$ and their corresponding association with the blocked and non-blocked paths well-known in the frame of percolation theory \cite{percolation2}.

\subsection{Modelling epidemics employing elementary percolation theory}
\begin{figure}[h!]
\centering
\includegraphics[clip,width=0.60\columnwidth]{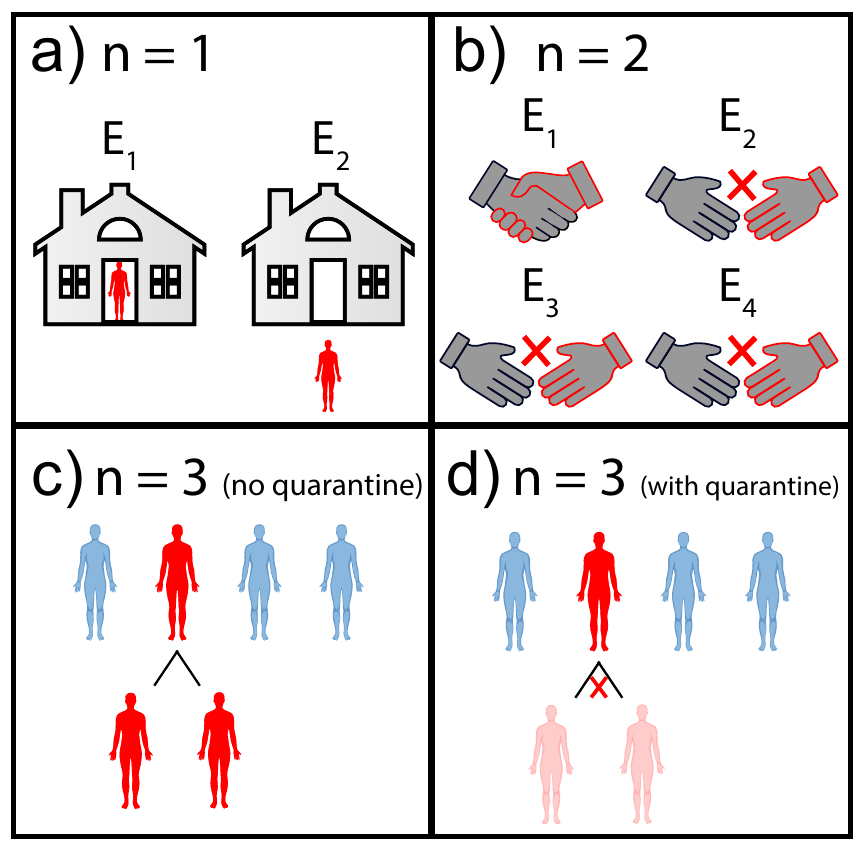}
\caption{\footnotesize  Schematic representation of the possibilities of infection by Covid-19 based on the number of connections $n$. a) For $n = 1$ there are two possible states $E_1$ and $E_2$, where $E_1$ represents a blocked path (infected person in quarantine) and $E_2$ a non-blocked path (infected person not in quarantine), but not in contact with other people. b) For $n = 2$ there are four possible states, labelled by $E_1$, $E_2$, $E_3$, and $E_4$. The interaction between an infected (hand outlined in red) and non-infected person (hand outlined in black) is represented by a handshake. Only state $E_1$ represents an actual percolation, i.e., infection since the contact interaction actually happened. c) For $n = 3$ one infected person (red color) can infect two others by not being on quarantine, while in d) the person respected the quarantine and avoided infecting two other people (represented by the light red color; blue colour indicates non-infected people). For $n = 3$, there are 8 combinatory possibilities, but for the sake of compactness we indicate only two of them.  Details in the main text. Figure generated using templates available in Ref.\,\cite{vecteezy}.}
\label{Fig-6}
\end{figure}

\begin{figure}[h!]
\centering
\includegraphics[clip,width=0.60\columnwidth]{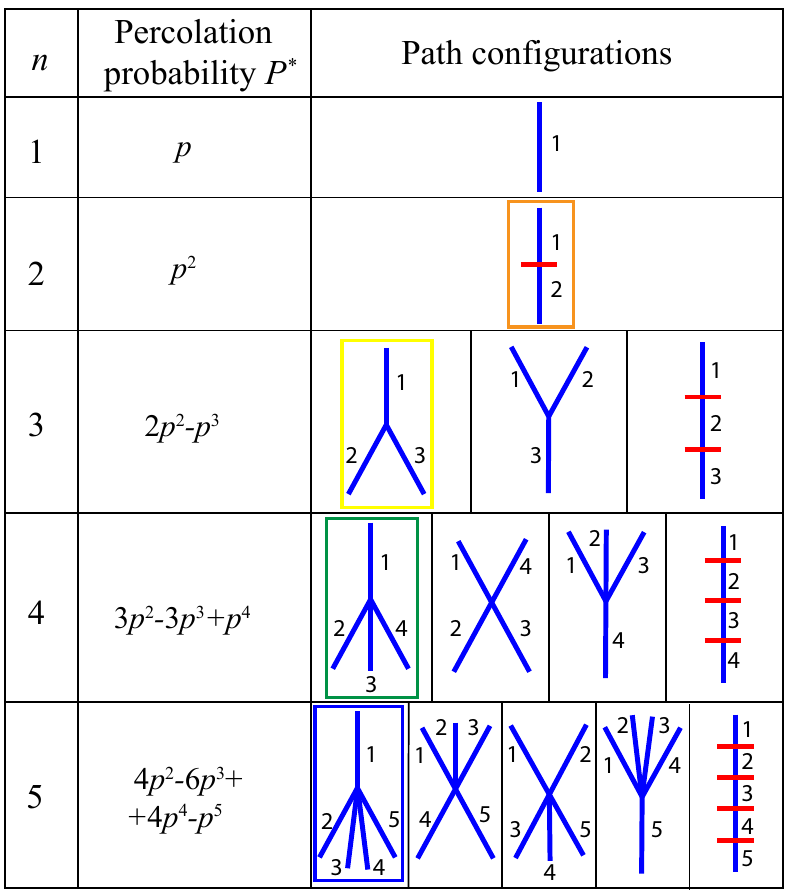}
\caption{\footnotesize
Number of connections $n$ (left column) with their corresponding percolation probability $P^*$ terms (middle column) and schematic representation of the possible path configurations (right column). The red lines represent a division of the blue paths. Each configuration has 2$^n$ possible states, considering all the possibilities of blocked and non-blocked paths. Details in the main text.}
\label{Fig-7}
\end{figure}

\begin{figure}[h!]
\centering
\includegraphics[clip,width=0.90\columnwidth]{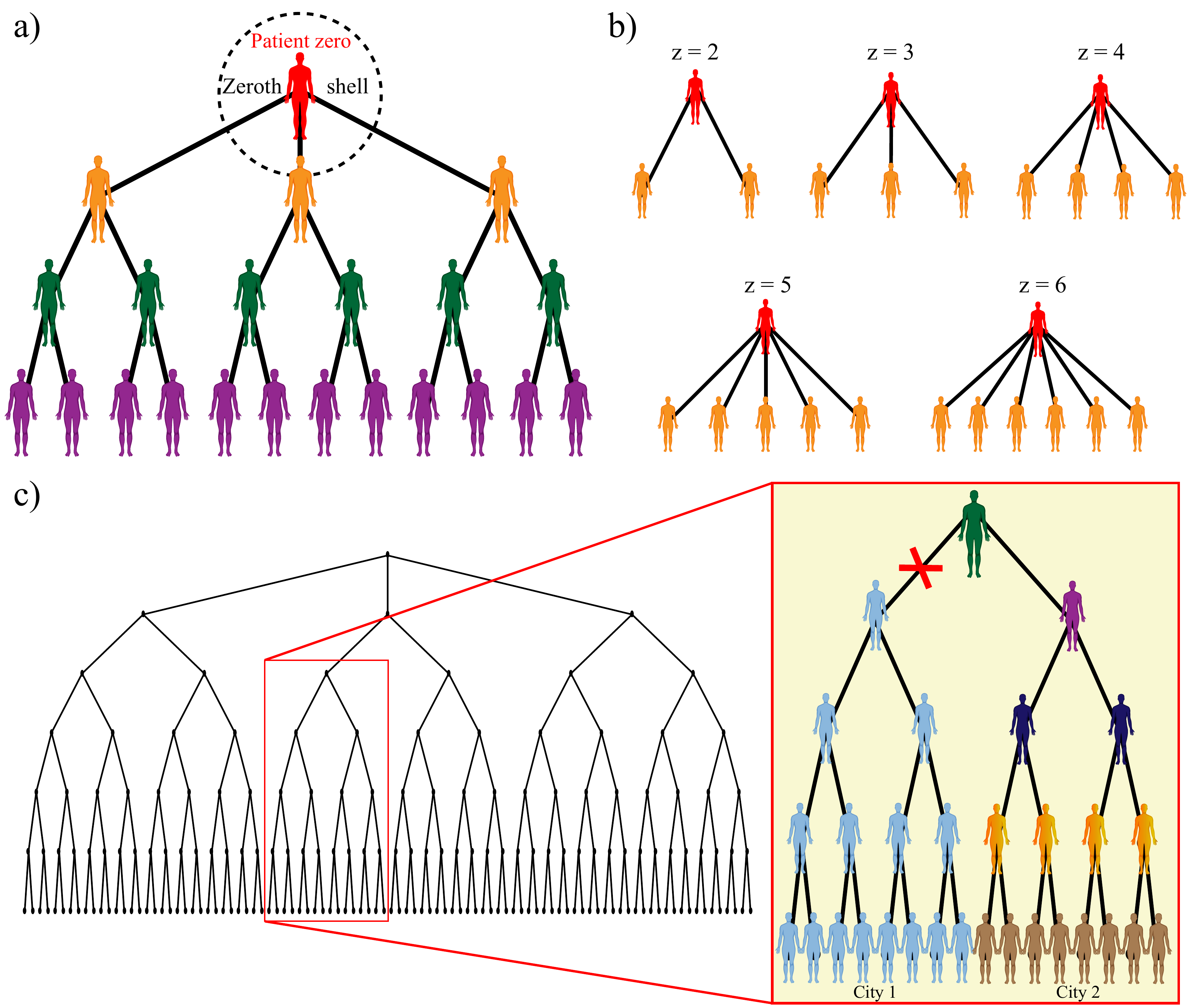}
\caption{\footnotesize a) Representation of the Bethe lattice for $z = 3$ \cite{vecteezy, Mathematica} where the zeroth shell is occupied by an infected person (in red color). In this configuration, each infected person can infect two other more. For the sake of compactness, we stick to only four shells in this representation of the Bethe lattice, namely the zeroth ($n' = 0$, red color), the first ($n' = 1$, orange color), the second ($n' = 2$, green color), and the third ($n' = 3$, purple color) shells. The starting point is represented by the patient zero (highlighted in red color). b) Representation of the number of people that an infected person (red color) \cite{vecteezy} can infect in terms of the number of nearest neighbors $z$ employing the Bethe lattice \cite{Mathematica} for $2 \leq z \leq 6$. c) Schematic representation of the Cayley tree for $z = 3$ for various shells. The zoomed region outlined in red depicts two distinct branches representing two distinct hypothetical cities 1 and 2. An infected person (green color) is unable to reach the city 1 and thus everyone there is healthy (blue color). On the right branch, the infected person can reach the city 2 and thus spreading the disease. The various colors employed to represent the people in the right branch represent the various subsequent shells.  More details in the main text.}
\label{Fig-8}
\end{figure}

In order to describe the probability of infection by Covid-19 or any other disease, we make use of elementary concepts in the frame of percolation theory \cite{Ricieri}, which determines the probability of percolation to take place in terms of non-blocked $p$ and  blocked $(1-p)$ connections. The probability of percolation is associated with the probability of a person being infected with Covid-19 through contact with an infected person (Fig.\,\ref{Fig-6}).
A blocked connection represents an infected person respecting the quarantine and thus ``blocking'' the spread of the disease. In the same way, the non-blocked connection refers to an infected person not joining the social distancing and thus, unfortunately,  contributing to the increased of probability of new people becoming infected. Furthermore, we make use of a simple 3 connections net to describe the Covid-19 spread, i.e., one infected person can infect two other people and so on (Fig.\,\ref{Fig-7}). Basically, for each number of connections $n$, we have a total probability $P_n$, which includes the percolation and non-percolation probabilities, cf.\,Fig.\,\ref{Fig-7}. In the simplest case, i.e.,  $n = 1$, there are only two possible states, namely the path is non-blocked $p$ or it is blocked $(1-p)$. Hence, the total probability in this case is given by $P_1 = p + (1-p) = [p + (1-p)]^1$. For $n = 2$, there are four possible states: connection 1 blocked and connection 2 non-blocked, connection 1 non-blocked and connection 2 blocked, connection 1 and 2 blocked, and connection 1 and 2 non-blocked, being the corresponding probabilities for each state given, respectively, by $2 \times (1-p)p$, $(1-p)(1-p)$, and $p^2$. In this case, the total probability is the sum of the probabilities of each state, given by $P_2 = 2 \times (1-p)p  + (1-p)(1-p) + p^2$, which can be rewritten as $P_2 = [p+(1-p)]^2$. Following such a logic, the generalized mathematical function of the total probability $P_n$, i.e., including the percolative and non-percolative probabilities, in terms of the number of connections $n$ reads \cite{Ricieri}:
\begin{equation}
P_n = [p + (1-p)]^{n} = 1.
\label{Pn}
\end{equation}
Note that $P_n$ includes both percolative and non-percolative probabilities, but we are interested only in the infection (percolative) probability. Thus, we consider only the terms of $P_n$ that contribute to the percolation probability. Considering $n = 3$, for instance,  we have:
\begin{equation}
P_3 = p^3 + 3(1-p)p^2 + 3(1-p)^{2}p + (1-p)^3 = 1.
\label{n3}
\end{equation}
However, the percolation probability, labelled here $P^*$, is given only by the terms $P^* = (2p^2 - p^3)$. Essentially, in order to determine the percolation probability, it is necessary to make a combinatory analysis employing the value of $n$ considering the probabilities for all possible states $E_n$ to exist. Then, we rule out all the probabilities contributions associated with the states presenting null chance of percolation to take place, making thus the percolation probability as the sum of all the probabilities of existence of all the remained states. Figure \ref{Fig-7} summarizes an extension of such an analysis for other values of $n$ by considering only the terms associated with the percolative probability.
At this point, we take into account the case for $n=3$ and also make a broader analysis considering the so-called Bethe lattice. As can be inferred from Fig.\,\ref{Fig-7}, for $n=3$ there is a path leading to two others, being the consecutive paths originating from each of these two not shown. By taking such consecutive paths into account, we can form infinite shells leading to the formation of a Bethe lattice, as shown in Fig.\,\ref{Fig-8}.
For the sake of completeness, we recall next the formation of such a lattice based on discussions reported in textbooks, such as in Ref.\,\cite{percolation2}. In the initial condition, a central site with a probability $p$ of being occupied and $(1-p)$ of being empty leads to an amount of $z$ paths (coordination number), which also has a site at the end of each path. Thus, we can make a direct analogy with the Ising-model for  integer spin $S$, where we have $S \pm 1$ configuring spin up or down, and $S = 0$ meaning that there is a vacancy in such a site. The set of sites formed by the central site and the $z$ paths constitute the zeroth shell of the lattice, cf.\,indicated in Fig.\,\ref{Fig-8} a). Then, each new site on the boundary of shell zero gives rise to $(z-1)$ new paths and sites forming shell number one. This can be performed continuously to form other shells until, at the end of the lattice, the last sites have only one path and there is no longer the possibility of forming new shells. In this context, taking the central site as reference so that we can follow a path from the central site to a site at the boundary of the final shell, it is necessary that there are both available paths and occupied sites, configuring a percolation. In this way, it is possible to consider the percolation probability $P^*$ as the probability of obtaining a path allowing the central site to be connected to a site in the final shell of the Bethe lattice. In an analogous way, we represent the non-percolation probability as $Q$. If each occupied site with probability $p$ leads to $(z-1)$ paths, we can estimate an average of $(z-1)p$ occupied sites. However, having in mind that a site can be occupied or not, the reduction of available paths at each shell will prevent to reach the final sites and therefore there is no percolation. Upon going from one shell to other if $(z-1)p < 1$, then the probability of reaching the final shell (percolation) is lowered.  Such analysis is important because it leads to the so-called percolation threshold $p_c=\frac{1}{(z-1)}$. Employing $z = 3$, for instance, we focus now on the factors that may prevent percolation. In this specific case, any site leads to two paths and each of them with a $Q$ probability of non-percolation. From the standard definitions of sets and probabilities, two events will be statistically independent if and only if the intersection between them can be written as a product of both probabilities \cite{Florescu}. Hence, the probability of non-percolation for $(z-1)$ will be $Q^2$. Also, the sites must be occupied for the percolation to occur, so that the total probability of non-percolation should be written as $Q = (1-p) + pQ^2$. The solutions of the latter are $Q = 1$, i.e., there is no percolation, and $Q = \frac{(1-p)}{p}$, which can be equal to zero if $p=1$ and thus percolation takes place. Analyzing the central site and the three paths originating from it, the percolation probability must take into account $p$ and discount the probability of non-percolation, i.e., $P^*=p-pQ^3$. Replacing here the previous solutions obtained for $Q$, we have two different cases. If $Q = 1 \Rightarrow P^* = 0$, which represents the non-percolative condition, i.e., $p < p_c$. If $Q = \frac{(1-p)}{p}$, then:
\begin{equation}
P^*=p\left\{1-\left[\frac{(1-p)}{p}\right]^3\right\},
\end{equation}
which has a phase transition order parameter-like behavior for the percolation. It is worth noting that such analysis is distinct for different values of $z$ since it is dependent on the number of possible paths. Also, note that $P^*$ is universal for $z = 3$ and does not depend on the length $S'$ of the path:
\begin{equation}
S' \propto \frac{1}{p-p_c},
\end{equation}
which is associated with the size of the lattice \cite{percolation2}. Analyzing the simplest case, i.e., a chain composed by a set of sites, it is possible to define the pair correlation function $G(r)$. The latter is associated with the percolation probability between two occupied reference sites separated by a distance $r$. Such a distance incorporates other sites that may be present between the occupied ones. The $G(r)$ function is given by $G(r) = e^{-r/\xi}$ \cite{percolation2}, where $\xi$ is the so-called correlation length $\xi = 1/(p_c - p)$. For $p \rightarrow p_c$, $\xi$ diverges and the $G(r) \rightarrow 1$. In the case of the Covid-19 spread, this would mean that if the number of non-blocked paths achieves such a threshold, i.e., a significant portion of the population does not join the social distancing, then a pronounced increase of the number of new infections over time would take place.  The very same mathematical treatment previously discussed can be employed upon considering the branches of the Cayley tree as cities, cf.\,Fig.\,\ref{Fig-8} c). It has been lately considered the effectiveness of the so-called intermittent quarantine \cite{robotdance}. The latter means that nearby cities would join the social distancing in pre-programmed different days. As a consequence, the economic activities can be retaken and  with relatively low probability of spreading the disease between such cities. The main idea behind lies on the fact that if an infected person is unable to reach one of the cities [Fig.\,\ref{Fig-8} c)], the probability of a person being infected in that specific city is lowered, as well as the probability of all subsequent infections that would take place.
In order to discuss the Cayley tree, we have used the concept of a regular tree, which means that the branches of the tree are constructed always in the same way employing a fixed $z$ number. However, random trees can also be employed \cite{Shlomo,Goltsev2008}, such as the Erd\H{o}s-R\'enyi network, where the construction of such tree is probabilistic and $z$ is not fixed. In the latter case, crossings between branches can take place. Furthermore, it is also reported in the literature the so-called pruning process in random tree \cite{Goltsev2015}, where each site, or vertex, is systematically removed over time. Initially, there are a few branches and over time the tree reaches a plateau, i.e., the number of branches is minimized and, upon continuing the pruning process, the tree itself disappears. Bringing this discussion to the case of the Covid-19 spread through the Cayley tree, we can associate each pruned vertex with either a person passing away or joining the social distancing [Fig.\,\ref{Fig-8} c)]. As an analogy with the random tree discussion, upon continuously pruning the tree, such as in the case of people joining the social distancing or passing away, the epidemic is faded away, i.e., the tree vanishes.
The effects of the selected quarantine can also be analyzed in the frame of the SEIR model, which also takes into account the number of people exposed to the disease. Thus, employing the usual factors for the SEIR model, the number of connected cities, particular regions, states or even countries and the number of people circulating from one region to another, it is possible to make a forecast of the number of  new infections \cite{robotdance}, cf.\,Fig.\,\ref{Fig-8} c).

\subsection{Switching on the interactions in the Bethe lattice}\label{Ising-Bethe}
The Ising-model can be applied to the Bethe lattice \cite{Baxter}. To this end, let us consider a lattice in which the central site has spin $\sigma_0$. The latter has $z = m_1$ neighbors, $m_2$ next-nearest neighbors, and so on until the last shell $n'$ which leads to $m_{n'}$, i.e., $n'$-th neighbors.
\begin{figure}[h!]
\centering
\includegraphics[clip,width=0.90\columnwidth]{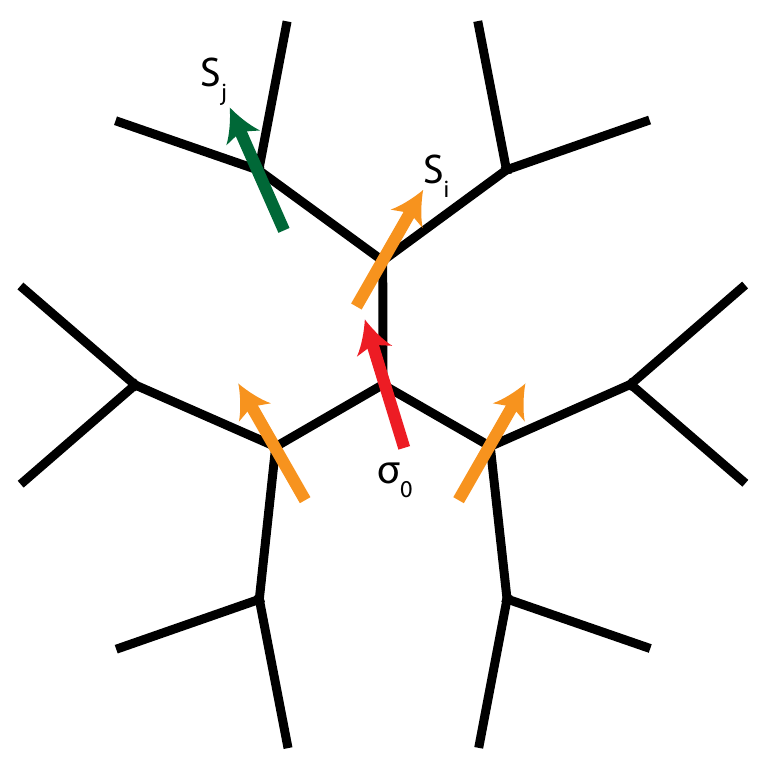}
\caption{\footnotesize Representation of spins distributed on the Bethe lattice \cite{Mathematica} with $z = 3$ showing a spin (red arrow) $\sigma_0$ at the central point and other neighbouring spins at sites $S_i$ (orange color arrow) and $S_j$ (green arrow). More details in the main text.}
\label{Fig-X}
\end{figure}
The total number of sites $C_{n'}$ on the lattice can be written as \cite{Baxter}:
\begin{equation}
C_{n'} = 1 + m_1 + m_2 + ... + m_{n'}.
\end{equation}
Furthermore, the number of sites in a network should increase with the number of shells \cite{percolation2}, as follows:
\begin{equation}
C_{n'} = (n')^d,
\end{equation}
where $d$ is the dimensionality of the lattice.
For a Bethe lattice, $C_{n'}$ reads \cite{Baxter}
\begin{equation}
C_{n'} = \frac{z[(z-1)^{n'}-1]}{z-2}.
\end{equation}
For an infinite number of shells, we have \cite{Baxter}
\begin{equation}
\lim_{n'\rightarrow\infty}\left\{ \frac{\ln C_{n'}}{\ln n'} \right\} = d\rightarrow\infty,
\end{equation}
that is, for a lattice with infinite shells we also have infinite dimensionality. For the sake of completeness, we present here a textbook-like discussion of the Ising-model in the Bethe lattice. Essentially, our goal is to make a connection between the Bethe lattice and the interaction parameter $\delta \varepsilon$. Before starting, we recall Eq.\,\ref{ising} where the Hamiltonian $H$ for the Ising-model considering longitudinal applied magnetic field is defined. The first step to calculate the physical quantities of interest is the \emph{construction} of the partition function \cite{Baxter}. The latter represents the sum over all possible accessible states (\emph{Zustandssumme}) of the system and reads:
\begin{equation}\label{partition}
Z = \sum e^{-\beta H}.
\end{equation}
In a magnetic system, the states are associated with the possible spin orientations. So, replacing Eq.\,\ref{ising} in \ref{partition} we have:
\begin{equation}
Z = \sum_S \exp\{\beta [J \sum_{i,j}S_iS_j + B\sum_i S_i]\}.
\label{probabilityP}
\end{equation}
Considering a particle in a reservoir in which the temperature is fixed and the particle is in a state $\psi_l$ with energy $E_l$, the number of remaining accessible states is determined by the multiplicity $\Omega$ calculated in terms of the difference between the total energy $E_{tot}$ and $E_l$. Thus, the probability that the particle occupies a state labelled by $\psi_l$ can be written as follows \cite{Ralph}:
\begin{equation}\label{probl}
P(\psi_l) = \emph{constant} \times \Omega(E_{tot}-E_l),
\end{equation}
where \emph{constant} refers to a normalization constant. The Boltzmann expression for the entropy, namely $S = k_B \ln[\Omega]$, allows us to calculate the entropy $S$ of a system as a function of $\Omega(E_{tot}-E_l)$. Considering that the reservoir is much larger than the
particle itself, i.e., there are many accessible states inside the reservoir, then $E_{tot}$ $>>$ $E_l$. Thus, since $E_l$ is very small compared to $E_{tot}$, the entropy difference $S(E_{tot} - E_l)$ can be expanded in a Taylor series \cite{Ralph}:
\begin{equation}
S(E_{tot}-E_l) = S(E_{tot}) + \frac{\partial S}{\partial E}\bigg|_{E=E_{tot}} \times (-E_l),
\end{equation}
and, since $\frac{1}{T} = \left(\frac{\partial S}{\partial E}\right)_{f.e.p.}$ \cite{Ralph},  where $E$ is the system's energy and f.e.p. refers to fixed external parameters, the entropy reads:
\begin{equation}
S(E_{tot}-E_l) = S(E_ {tot}) - \frac{1}{T} E_l.
\end{equation}
Thus, $\Omega$ can be written as:
\begin{equation}
\Omega = \exp \left[\frac{S(E_{tot}-E_l)}{k_B} \right] = \exp \left[\frac{S(E_{tot})-\frac{E_l}{T}}{k_B} \right].
\label{multiplicity}
\end{equation}
Eq.\,\ref{multiplicity} can be replaced into Eq.\,\ref{probl} to determine the probability of the particle to be found at  $\psi_l$ state:
\begin{equation}\label{probentropy}
P(\psi_l) = constant \times \exp \left[\frac{S(E_ {tot})-\frac{E_l}{T}}{k_B} \right].
\end{equation}
Since the total energy is fixed constant, we now write that $constant \times \exp{[S(E_{tot})/k_B]} = constant'$ and thus:
\begin{equation}
P(\psi_l) = constant' \times \exp \left[\frac{-E_l}{k_B T} \right].
\end{equation}
Furthermore, the total probability is defined as $\sum_l P(\psi_l) = 1$, which allows us to calculate the normalization constant and  thus achieve the following expression:
\begin{equation}
P(\psi_l) = \frac{\exp \left[\frac{-E_l}{k_B T}\right]}{\sum_{l'} \exp \left[\frac{-E_{l'}}{k_B T} \right]}.
\label{ppsi}
\end{equation}
Note that the term $e^{- E_l/k_BT}$ represents only the particular state $\psi_l$ and,  in the denominator term the sum is over all accessible states. Thus, it is needed to use the index $l'$ to differ such terms. Equation\,\ref{ppsi} can be applied to the Bethe lattice taking into account the possible $S_i, S_j$ states \cite{Baxter}. Hence, we rewrite Eq.\,\ref{partition} as a function of a non-normalized probability distribution $P(S)$ as:
\begin{equation}
Z = \sum_S P(S),
\end{equation}
where, by comparison, $P(S)$ is the term of the sum in Eq.\,\ref{probabilityP}. This is a reasonable association, since the  non-normalized probability distribution function  $P(S)$ can be written in terms of the multiplicity $\Omega$, which in turn is associated with the probability that a particle occupies a certain state. Thus, we have:
\begin{equation}
P(S) = \exp \left\{\beta \left[J \sum_{i,j}S_iS_j + B\sum_i S_i \right] \right\}.
\end{equation}
Now, we focus on the zeroth shell of the Bethe lattice.
Equation\,\ref{probabilityP} is valid for a chain of spins, i.e., a one-dimensional case. However, in order to analyse a bidimensional case, for instance, the Bethe lattice can be employed. It is necessary to take into account interaction between the spin at the central point, labelled by $\sigma_0$, and the spins $S_i$ at each site of the $z$ paths connected to this central point. As previously mentioned, for statistically non-correlated events, the total probability will be given by the product of the individual probabilities of each event separately. Hence, the total probability of the system to be at a state $S$ reads \cite{Baxter}:
\begin{equation}
P(S) = \exp\{\beta [J\sum_{i=1} S_i \sigma_0 + B\sigma_0]\} \prod_{j=1}^z Q_{n'}(S^{(j)}),
\label{PS}
\end{equation}
where,
\begin{equation}
Q_{n'}[S^{(j)}] = \exp\{\beta[J\sum_{i,j}S_iS_j + B\sum_i S_i ]\}.
\label{Qn}
\end{equation}
Equation \ref{PS} gives the probability $P(S)$ in terms of the interaction between spin $\sigma_0$ at the central point and each of its first neighbors at a site $i$, cf.\,sketched in Fig.\,\ref{Fig-X}. Then, all subsequent interactions between the spin at a site $i$ and its nearest neighbors at a site $j$ are considered, which is represented by the $Q_{n'}[S^{(j)}]$ product. Note that the interaction between $S_i$ spins are not taken into account, since such paths in the Bether lattice are not connected. As discussed in Section \ref{Section2}, Eq.\,\ref{contamination} incorporates, when taking into account infected and non-infected people, the $\delta\varepsilon$ \textit{interaction}, which is equivalent to the magnetic exchange coupling constant $J$ in the Ising-model. Such an equivalence enables us to calculate the probability $P(p)$ of a person being infected or not taking into account $\delta\varepsilon$ and the interactions between the person at the central site $p_0$ and its first neighbors $p_i$, as well as the interaction between a person labelled by $p_i$ and its nearest neighbor $p_j$ (Fig.\,\ref{Fig-X}):
\begin{eqnarray}
P(p)&=&\exp\bigg{\{}-C_h+4\delta\varepsilon\sum_{i=1} \left[\left(\frac{1}{2}+p_i\right)\left(\frac{1}{2}-p_0\right) + \right. \nonumber \\
&& \left.+\left(\frac{1}{2}-p_i\right)\left(\frac{1}{2}+p_0\right)\right]p_ip_0\bigg{\}} \prod_j^3Q_{n'}(p^{(j)})
\label{contador1}
\end{eqnarray}
where,
\begin{eqnarray}
Q_{n'}(p)&=&\exp\bigg{\{}-C_h+4\delta\varepsilon\sum_{i\neq j=1}^N \left[\left(\frac{1}{2}+p_i\right)\left(\frac{1}{2}-p_j\right) + \right. \nonumber \\
&& \left.+\left(\frac{1}{2}-p_i\right)\left(\frac{1}{2}+p_j\right)\right]p_ip_j\bigg{\}},
\label{contador2}
\end{eqnarray}
and therefore, $P(p) \propto e^{-\delta\varepsilon}$. The exponential decay of $P(p)$ in terms of the increase of the interacting parameter $\delta\varepsilon$ means that, upon increasing $\delta\varepsilon$, the probability of a person being infected is reduced. In other words, an increase of $\delta\varepsilon$ reflects on an increase in the number of infected people, decreasing thus the number of non-infected people that can be infected.
Note that, for the case of two infected people $p_i = p_j = +1/2$ or two non-infected people $p_i = p_j = -1/2$ interacting with each other, the sums in Equations \ref{contador1} and \ref{contador2} are null. Indeed, such sums are non-zero only for the case of an interaction between a non-infected person and an infected person (or vice-versa). Thus, we have the probability of the disease spread in terms of the interactions $\delta\varepsilon$ between neighboring people in a similar way as in the previously discussed case for the Ising-model on the Bethe lattice.

\subsection{The Bethe lattice and the spread of Covid-19}
Bringing this whole discussion to the context of the Covid-19 spread, we can associate an occupied site in the Bethe lattice with an infected person and the paths with the displacement of such people and their possibility of infecting healthy people (Fig.\,\ref{Fig-8}). If an infected person \emph{travels} to the end of a path and finds a healthy person, that person becomes infected and can go on infecting two more others, for example. However, if a healthy person remains in quarantine and it is not in the infection path, the site becomes empty reducing the probability of percolation. In the same way, if an infected person respects the quarantine, the path leading to the contamination is interrupted and the percolation probability is also reduced. It is clear that upon increasing the number of possible paths the data operations turn into a complex exercise, preventing thus the resolution of the problem \emph{by hand}. Hence, aiming to deal with such a complexity, advanced data analysis techniques are required. This includes, for instance, Markov Chain Monte Carlo methods such as the emcee package, and Machine Learning based algorithms such as the ones presented in the SciKitLearn package \cite{Big-data,Lee2018,Drake2016,Barber2011,Goodman2013,Pedregosa2011}. Several tracking methods have been employed in order to collect big-data sets of millions of internet users aiming to make a proper mathematical description of a collective behavior, such as epidemic outbreaks. Among such methods, the ARGO (AutoRegression with GOogle search data) stands out, being applied, for instance, for the case of the influenza epidemics \cite{Kou2015}. Recently, the ARGO method was employed in the case of Covid-19 to make a real-time forecast about the disease spread in small provinces in China \cite{Santillana2020}. As discussed previously, the behavior of the percolation probability $P^*$ as a function of $p$ is altered when the number of connections $n$ is increased [Fig.\,\ref{Fig-9}]. For a system presenting a relatively high number of connections $n$, a percolation probability distribution takes place.
This is the case, for instance, of crowded places, as in the so-called \emph{favelas} in Brazil, where each person represents a connection $n$ that may contribute to the increase of the percolation probability distribution, i.e., the dissemination of the disease. In order to analyze the behavior of such distribution as a function of $p$ for various values of $n$, it is natural to employ a distribution function. Indeed, note that the percolation probability polynomial function (Fig.\,\ref{Fig-7}) can be approximated by:
\begin{equation}
P^*(n, p) = \frac{1}{2} + \frac{1}{\pi}\arctan\left[n\frac{(p - p_c)}{p_c}\right],
\label{vladi}
\end{equation}
where $p_c$ refers to the critical value of connections. Equation \,\ref{vladi} is also called the Cauchy distribution function \cite{Morio}.
 Equations.\,\ref{vladi} and \ref{distributionfunction} represent a probability distribution, being such mathematical functions asymptotically different so that Eq.\,\ref{vladi} cannot be rewritten in the form of Eq.\,\ref{distributionfunction} and vice-versa.

\subsection{Data analysis and discussion in the frame of the percolation theory}
\begin{figure}[h!]
\centering
\includegraphics[clip,width=0.66\columnwidth]{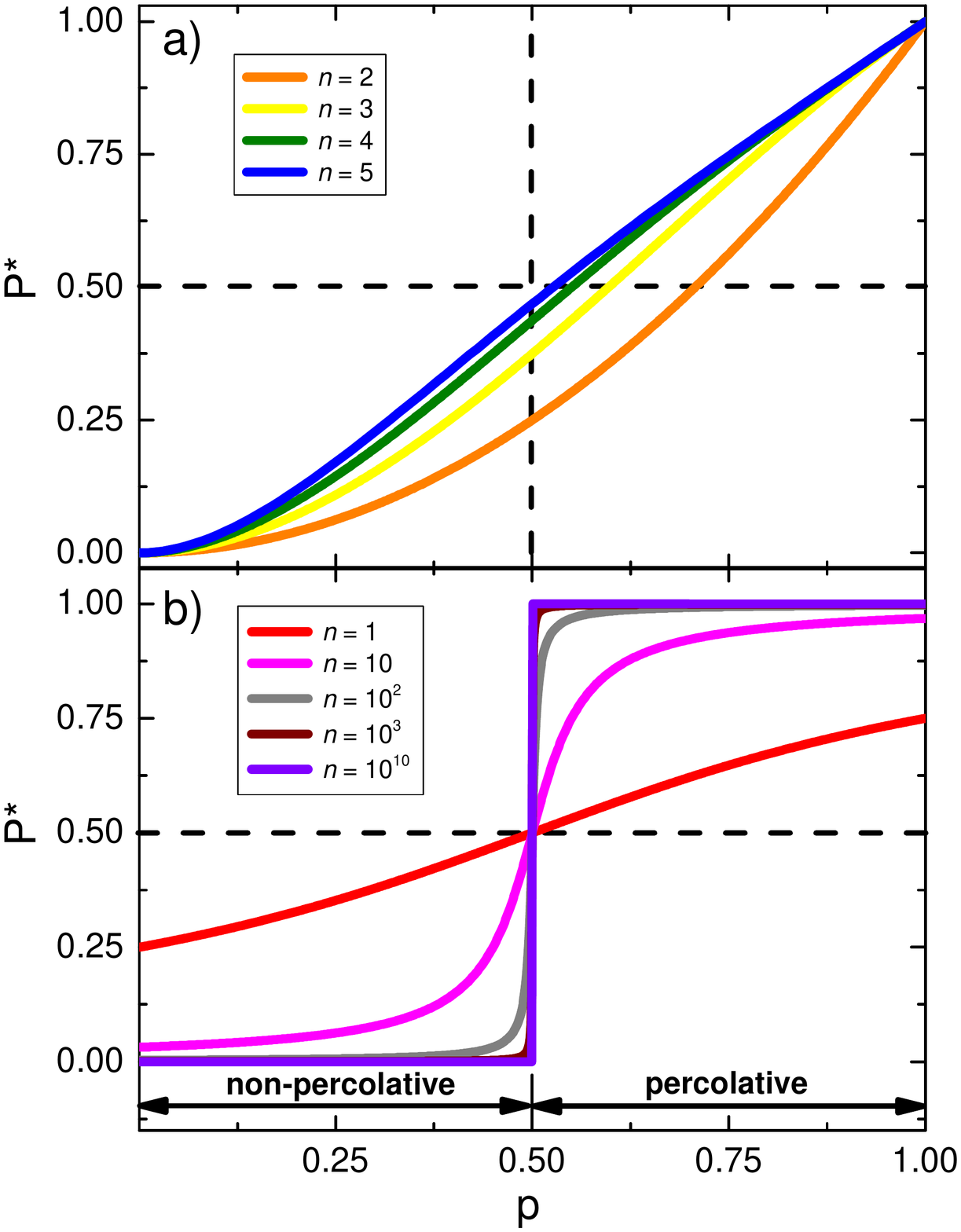}
\caption{\footnotesize  Percolation probability $P^*$ as a function of the non-blocked connections $p$ for various number of net connections $n$ employing a) the percolation probabilities polynomials shown in Fig.\,\ref{Fig-7} and b) Eq.\,\ref{vladi}. Details in the main text.}
\label{Fig-9}
\end{figure}

Making use of the adaptation of the percolation theory for the Covid-19 spread previously discussed, we discuss now the effects on the percolation probability upon increasing the number of connections $n$, cf.\,shown in Fig.\,\ref{Fig-9}.
Upon comparing, for instance, the behavior between the percolation probability for $n = 2$ and 5, we observe that $P^*$ grows more rapidly for higher values of $n$, indicating that the more the number of connections $n$, the faster the percolation probability increases in terms of the non-blocked paths $p$. For finite values of $n$, the percolation probability is enhanced upon increasing the number of non-blocked paths $p$, i.e., the probability of people getting infected by Covid-19 increases proportionally to the number of people not in quarantine. However, when $n \rightarrow \infty$ the percolation probability approaches 1 for $p \rightarrow 1/2$, i.e., if people all around the world would interrupt the social distancing, the probability of people getting infected by Covid-19 would be enhanced dramatically. In fact, the more people get infected, more and more number of connections $n$ takes place, decreasing thus the minimal number of non-blocked paths $p$ (people not respecting the quarantine) for the percolation, i.e., spread of the disease, to occur. In other words, the probability of people getting infected by Covid-19 in terms of the people not respecting the quarantine grows faster when the total number of connections (infections) $n$ is increased. Yet, an interesting analogy of such percolation probability can be performed. Since the fraction of people in quarantine and non-infected people is linked to each other, i.e., as one increases the other decreases so that $P^* + Q$ = 1, one can make an analogy with the Physics of semiconductors and the well-established law of mass action \cite{Kittel}. The latter establishes that the product of the density of electrons and holes at a certain temperature is constant and it also depends on the product of their corresponding masses. In other words, the amount of electrons promoted from the valence to the conduction band is equal to the number of remaining holes in the valence band. As an analogy, $p$ and $(1-p)$ can be interpreted in the same way since any increase (decrease) in $p$ implies a decrease (increase) in exactly the same proportion in $(1-p)$.

\section{What's next?}\label{WN}
Besides the need of a more appropriate mathematical approach in order to describe properly the epidemic curves of a particular disease, there are other inherent factors of our society that can be developed in order to minimize the social impact of other epidemics  in the future. These include the record and divulgation of reliable data sets in the frame of epidemics by health agencies in order to avoid or properly brake the outbreak of a disease. The development and design of mobile apps  \cite{Bulchandani2020} can serve as an important tool to be employed in a more efficient social distancing. Such mobile apps can indeed be very useful in the management of the disease spread. However, their use collide with ethical aspects regarding the control of each individual's information and locations by the government, which in many cases can preclude the application of such apps. As pointed out in Ref.\,\cite{Heesterbeek2015}, the record of the epidemic data must be performed and analyzed together with other relevant data sources, such as demographic, genetic, and travel patterns for the various locations and temporal scales aiming the prevention and the containment of epidemics. Furthermore, it is highly required that the health agencies make a proper planning about the construction of hospital beds in the case of an epidemic, being obviously evident that the construction speed of such hospital beds in different countries will be distinct, impacting thus on the treatment of infected people. Even though the coronavirus was already known by the scientific community a few years ago \cite{Zhu2018} the outbreak of the disease could not be avoided. Hence, it  becomes clear that the discovery of a new virus should be accompanied, if possible, by the discovery of the antidotes and or vaccines. This points to the urge of high-level scientific research focussed on the prevention or management of future viral epidemics. Also, the incorporation of discussions on the risk and possible consequences of epidemics in the curriculum of primary and high schools may play an important role on the resulting awareness of younger people regarding profilatic measurements which could, indeed, reduce the impact of future epidemics.

\section{Conclusions}
We have reviewed the key-related aspects to the SIR model for various epidemics and provided a brief discussion about its mathematical description. An adaptation of a $S = 1/2$ Ising-like model was presented and we deduced the logistic function, usually employed to describe epidemic curves among other phenomena. We have shown that the temporal evolution of the number of infections and fatalities, described by the logistic function, has some resemblance with a distorted Fermi-Dirac-like distribution function found in the celebrated Landau Fermi-liquid theory. Our analysis demonstrated that a Gaussian-type function suffices to describe the epidemic curves and that the quarantine plays a crucial role in the amendment of Covid-19 spread. % in terms of the number of connections $n$ and the number of blocked (people on quarantine) and non-blocked (people not on quarantine) paths, being associated with the probability of percolation (infection).
The fundamental concepts of the Cayley tree and the Bethe lattice were discussed and a connection with the Ising-model was made. Yet, we have demonstrated that the percolation (infection) probability in terms of the number of people not respecting the social distancing sets in more rapidly when the total number of cases is increased, making thus evident the importance of the quarantine in the suppression of Covid-19 spread. We hope that the governmental health agencies can have benefits from it. The present work is, at some extent, an appeal to world leaderships to adopt the social distancing to brake the spread of Covid-19 before a vaccine is discovered and released. Such social distancing is crucial since there are possible sub-notifications of the number of new infection cases that could mask an even worse scenario of the number of infections. Our appeal is relevant in the attempt to prevent a second wave or even small outbreaks of the number of infections. This is particularly true in countries, such as, for instance, India and some African countries, for which the epidemic curve is still ascending up to date. The mass media plays an important role regarding the dissemination of reliable information aiming to properly aware the population, which can also be considered, in some cases, as an external factor that can indirectly influence the shape of the epidemic curves. Furthermore, we have reviewed the basics of percolation theory and employed it to describe the Covid-19 spread. The analytic solutions for the mathematical models, presented here, can serve as a basis for more complex cases involving numerical calculations. We have provided several distinct paths that can be employed to describe the epidemic curves for any disease. Such elementary discussions presented here can be useful in the application to other collective phenomena.  As discussed by the authors of Ref.\,\cite{Vespignani2020}, it is not straightforward to make a forecast of the epidemic curves  since the variables associated with the epidemic might change over time. Indeed, as pointed out in Ref.\,\cite{Bedford2019}, the outbreak of epidemics should be faced treated in a much broader context. The knowledge from various other research areas, besides epidemiology, can play a crucial role in a deeper discussion about the epidemics prevention, such as logistics and crisis management \cite{Bedford2019}. The defiance of the twenty-first century, e.g., poverty, climate change, and urbanization  significantly contribute to epidemics to happen more likely \cite{Bedford2019}. Also, it is crucial to learn from the negative outcomes of past epidemics, such as the Ebola outbreak in Congo \cite{Gbalo2019}, aiming to prevent other possible future epidemics. Furthermore, several countries in the world are experiencing the epidemic scenario for the first time, being difficult for such countries to properly manage the situation. Thus, a more embracing and modern approach on the containment of epidemics must be developed.  Also, it is mandatory the inclusion of strategic plans in future public health governmental policies. Last but not least, this work was written during our period of quarantine.

\begin{backmatter}

\section*{Declarations}
\section*{Competing interests}
The authors declare that they have no competing interests.

\section*{Author's contributions}
IFM, LS, and GOG carried out the calculations and generated the figures. IFM, LS, and MdeS wrote the paper with contributions from GOG and ACS. All authors read and approved the final manuscript. M.de.S. conceived and supervised the project.

\section*{Acknowledgements}
  MdeS acknowledges financial support from the S\~ao Paulo Research Foundation -- Fapesp (Grants No. 2011/22050-4 and 2017/07845-7), National Council of Technological and Scientific Development -- CNPq (Grants No.\,302498/2017-6).
ACS acknowledges CNPq (Grant No.\,305668/2018-8). This work was partially granted by Coordena\c c\~ao de Aperfei\c coamento de Pessoal de N\'ivel Superior - Brazil (Capes) - Finance Code 001 (Ph.D. fellowship of IFM and LS).
%%%%%%%%%%%%%%%%%%%%%%%%%%%%%%%%%%%%%%%%%%%%%%%%%%%%%%%%%%%%%
%%                  The Bibliography                       %%
%%                                                         %%
%%  Bmc_mathpys.bst  will be used to                       %%
%%  create a .BBL file for submission.                     %%
%%  After submission of the .TEX file,                     %%
%%  you will be prompted to submit your .BBL file.         %%
%%                                                         %%
%%                                                         %%
%%  Note that the displayed Bibliography will not          %%
%%  necessarily be rendered by Latex exactly as specified  %%
%%  in the online Instructions for Authors.                %%
%%                                                         %%
%%%%%%%%%%%%%%%%%%%%%%%%%%%%%%%%%%%%%%%%%%%%%%%%%%%%%%%%%%%%%

% if your bibliography is in bibtex format, use those commands:
%\bibliographystyle{bmc-mathphys} % Style BST file (bmc-mathphys, vancouver, spbasic).
%\bibliography{bmc_article}      % Bibliography file (usually '*.bib' )

\begin{thebibliography}{10}
\bibitem{Chinazzi} Chinazzi M, Davis JT, Ajelli M et al (2020) The effect of travel restrictions on the spread of the 2019 novel coronavirus (COVID-19) outbreak. Science 368(6489):395-400.
\bibitem{Servick} Servick K, Cho A, Couzin-Frankel J et al (2020) Coronavirus disruptions reverberate through research. Science 367(6484):1289-1290.
\bibitem{Cohen} Cohen J, Kupferschmidt K (2020) Countries test tactics in ‘war’ against COVID-19. Science 367(6484):1287-1288.
\bibitem{Rzymski} Rzymski P, Nowicki M (2020) Preventing COVID-19 prejudice in academia. Science 367(6484):1313.
\bibitem{Liu} Liu K (2020) How I faced my coronavirus anxiety. Science 367(6484)1398.
\bibitem{Vinko} Zlati\'c V, Barja\u{s}i\'c I, Kadovi\'c A et al (2020) Bi-stability of SUDR+K model of epidemics and test kits applied to COVID-19. arXiv preprint. arXiv:2003.08479v2.
\bibitem{Squazzoni2020} Squazzoni F, Polhill JG, Edmonds B et al (2020) Computational models that matter during a global pandemic outbreak: a call to action. J Art Soc Soc Sim 23(2):10 (2020).
\bibitem{Mallapaty} Mallapaty S (2020) Why does the coronavirus spread so easily between people? Nature 579:183.
\bibitem{Foppa} Foppa IM (2016) A Historical Introduction to Mathematical Modeling of Infectious Diseases: Seminal Papers in Epidemiology. Academic Press, San Diego.
\bibitem{Kramer} Kr\"amer A, Kretzschmar M, Krickeberg K (2010) Modern infectious disease epidemiology: concepts, methods, mathematical models, and public health. Springer Science \& Business Media, New York.
\bibitem{Giesecke} Giesecke J (2017) Modern Infectious Disease Epidemiology. CRC Press, Boca Raton.
\bibitem{White} Vynnycky E, White RG (2010) An introduction to infectious disease modelling. Oxford University Press, Oxford.
\bibitem{Shlomo} Havlin S, Cohen R (2010) Complex networks: structure, robustness and function. Cambridge University Press, Cambridge.
\bibitem{Nelson2014} Nelson KE, Williams CM (2014) Infectious disease epidemiology: theory and practice. Jones \& Bartlett Publishers, Burlington.
\bibitem{Frauenthal2012} Frauenthal JC (2012) Mathematical modeling in epidemiology. Springer Science \& Business Media.
\bibitem{Ma2009} Ma S, Xia Y (2009) Mathematical understanding of infectious disease dynamics. World Scientific, Singapore.
\bibitem{Brauer2008} Brauer F, Driessche PD, Wu J (2008) Lecture notes in mathematical epidemiology. Springer, Berlin.
\bibitem{Chen2020} Chen YC, Lu PE, Chang CS et al (2020) A time-dependent SIR model for COVID-19 with undetectable infected persons. arXiv preprint. arXiv:2003.00122.
\bibitem{Dehning2020} Dehning J, Zierenberg J, Sptizner FP et al (2020) Inferring change points in the spread of COVID-19 reveals the effectiveness of interventions. Science. doi:10.1126/science.abb9789.
\bibitem{R0covid19} Sanche S, Lin YT, Xu C et al (2020) High contagiousness and rapid spread of severe acute respiratory syndrome coronavirus 2. Emerg Infect Dis. doi:10.3201/eid2607.200282.
\bibitem{R0SARS} Wallinga J, Teunis P (2004) Different epidemic curves for severe acute respiratory syndrome reveal similar impacts of control measures. Amer J Epid 160(6):509-516 (2004).
\bibitem{R0ebola} Khan A, Naveed M, Dur-e-Ahmad M et al (2015) Estimating the basic reproductive ratio for the Ebola outbreak in Liberia and Sierra Leone. Inf Dis Pov 4(13). doi:10.1186/s40249-015-0043.
\bibitem{Xiao2020} Li H, Xiao H, Zhu R et al (2020) Warmer weather and global trends in the coronavirus COVID-19. medRxiv preprint. doi:10.1101/2020.04.28.20084004.
\bibitem{Julia2020} Gog JR (2020) How you can help with COVID-19 modelling. Nat Rev Phys 2:274-275.
\bibitem{Hethcote2000} Hethcote HW (2000) The mathematics of infectious diseases. SIAM Review 42(4):599-653.
\bibitem{Grassly} Grassly NC, Fraser C (2008) Mathematical models of infectious disease transmission. Nat Rev Micr 6:477-487.
\bibitem{Ising} Ising E (1925) Beitrag zur Theorie des Ferromagnetismus. Zeitschrift f\"ur Physik 31:253-258.
\bibitem{SR} Gomes G, Stanley HE, de Souza M (2019) Enhanced Grüneisen parameter in supercooled water. Sci Rep. doi: 10.1038/s41598-019-48353-4.
\bibitem{Barto} Bartosch L, de Souza M, Lang M (2010) Scaling theory of the Mott transition and breakdown of the Grüneisen scaling near a finite-temperature critical end point. Phys Rev Lett 104(24):245701.
\bibitem{2015} de Souza M, Bartosch L (2015) Probing the Mott physics in $\kappa$-(BEDT-TTF)$_2$X salts via thermal expansion. J Phys: Condens Matter 27(5):053203.
\bibitem{PRL2020} Mello IF, Squillante L, Gomes GO et al (2020) Griffiths-like phase close to the Mott transition. arXiv preprint. arXiv:2003.11866.
\bibitem{PRB2019} Gomes GO, Squillante L, Seridonio AC et al (2019) Magnetic Gr\"uneisen parameter for model systems. Phys Rev B 100(5):054446.
\bibitem{econophysics} Schinckus C (2018) Ising model, econophysics and analogies. Physica A 508:95-103.
\bibitem{Bar-Yam2020} Siegenfeld AF, Bar-Yam Y (2020) Negative representation and instability in democratic elections. Nat Phys 16:186-190.
\bibitem{Jordan2020} Jordan S (2020) From ferromagnets to electoral instability. Nat Phys 16:125-126.
\bibitem{Pinol2012} Crisostomo CP, Pi\~nol CMN (2012) An Ising-based model for the spread of infection. Int Schol Sci Res \& Inn. 6(7):735-737.
\bibitem{McKendrick1927} Kermack WO, McKendrick AG (1927) A contribution to the mathematical theory of epidemics. Proc R Soc Lond Series A 115(772):700–721.
\bibitem{Amaku2019} Lara CGA, Massad E, Lopez LF et al (2019) Analogy between the formulation of Ising-Glauber model and Si epidemiological model. J Appl Math and Phys 7(5):1052-1066.
\bibitem{Cushing2008} Chitnis N, Hyman JM, Cushing JM (2008) Determining important parameters in the spread of malaria through the sensitivity analysis of a mathematical model. Bull Math Biol 70(5):1272-1296.
\bibitem{Baxter} Baxter RJ (1982) Exactly solved models in statistical mechanics. Dover, Mineola.
\bibitem{Nolting} Nolting W, Ramakanth A (2009) Quantum theory of magnetism. Springer, Berlin.
\bibitem{Brauer2012} Brauer F, Castillo-Chavez C (2012) Mathematical models in population biology and epidemiology. Springer, New York.
\bibitem{nyt} Brazil coronavirus map and case count. https://www.nytimes.com/interactive/2020/world/americas/brazil-coronavirus-cases.html.
\bibitem{bbc} Coronavirus: hospitals in Brazil's São Paulo 'near collapse'. https://www.bbc.com/news/world-latin-america-52701524.
\bibitem{Bulchandani2020} Bulchandani VB, Shivam S, Moudgalya S et al (2020) Digital herd immunity and COVID-19. arXiv preprint. arXiv:2004.07237.
\bibitem{2019} Cerdeiri\~na CA, Troncoso J, Gonz\'alez-Salgado D et al (2019) Water’s two-critical-point scenario in the Ising paradigm. J Chem Phys 150(24):244509.
\bibitem{Gene} Stanley HE (1971) Introduction to phases transitions and critical phenomena. Oxford Science Publications, New York.
%\bibitem{MW} N. D. Mermin and H. Wagner, Phys. Rev. Lett. \textbf{17}, 1133 (1966).
%\bibitem{transverse} J. Wu, L. Zhu, Q. Si, J. Phys. Conf. Ser. \textbf{273}, 012019 (2011).
\bibitem{Hollingsworth2020} Anderson RM, Heesterbeek H, Klinkenberg D et al (2020) How will country-based mitigation measures influence the course of the COVID-19 epidemic? The Lancet 395(10228):931-934.
\bibitem{Earn2014} Ma J, Dushoff J, Bolker BM et al (2014) Estimating initial epidemic growth rates. Bull Math Biol 76(1):245-260.
\bibitem{Mackey} Foreman-Mackey D,  Agol E, Ambikasaran S,  Angus R (2017) Fast and Scalable Gaussian Process Modeling with Applications to
Astronomical Time Series. The Astronomical Journal, 154:220 (21pp).
\bibitem{Stewart} Stewart J (2011) Calculus – early transcendentals. Cengage Learning, Belmont.
\bibitem{Ricieri} Lecture Notes of Prof. Aguinaldo P. Ricieri, Curso Prandiano, Anglo Tamandar\'e - S\~ao Paulo (1996).
\bibitem{Kittel} Kittel C (2005) Introduction to solid state physics vol. 8. John Wiley \& Sons, Hodoken.
\bibitem{kinetics} Burnham AK (2017) Use and misuse of logistic equations for modeling chemical kinetics. J Therm Anal Calor 127:1107-1116.
\bibitem{anaerobic} Donoso-Bravo A, P\'erez-Elvira SI, Fdz-Polanco F (2010) Application of simplified models for anaerobic biodegradability tests. Evaluation of pre-treatment processes. Chem Eng J 160(2):607-614.
\bibitem{germination} Schimpf DJ, Flint SD, Palmblad IG (1977) Representation of germination curves with the logistic function. Ann of Bot 41(6):1357-1360.
\bibitem{Kyurkchiev} Kyurkchiev N, Markov S (2016) On the Hausdorff distance between the Heaviside step function and Verhulst logistic function. J of Math Chem 54:109-119.
\bibitem{worldometers} COVID-19 coronavirus pandemic. https://www.worldometers.info/coronavirus/.
\bibitem{ebola} Sarukhan A (2016) Ebola: two years and 11,300 deaths later. https://www.isglobal.org/en/ebola.
\bibitem{SARS} Forecasting cases \& duration of severe acute respiratory syndrome (SARS). https://condellpark.com/kd/sars.htm.
\bibitem{H1N1} Gawryszewski VP, Neumann AILCP, Sesso RCC et al (2009) Trend and profile of non communicable diseases in the state of S\~ao Paulo. Bol. epidemiol. paul. 6(66):4-16.
\bibitem{Media} Zafar A (2020) 'They have changed the course of this outbreak:' Revelations from handling of coronavirus in China. https://www.cbc.ca/news/health/covid-19-china-epicurve-1.5479983.
\bibitem{fermiliquid} Varma CM, Nussinov Z, van Saarloos W (2002) Singular or non-Fermi liquids. Phys Rep 361(5):267-417.
\bibitem{Pines} Pines D (2018) Theory of quantum liquids: normal Fermi liquids vol. 1. CRC Press, Boca Raton.
\bibitem{Landau1987} Landau LD, Lifshitz EM, Pitaevskij LP (1987) Statistical physics, part 2: theory of the condensed state. Course of theoretical Physics, vol. 9. Butterworth-Heinenann, Oxford.
\bibitem{CDC} Centers for Diseases Control and Prevention (2020) People who are at higher risk for severe illness. https://www.cdc.gov/coronavirus/2019-ncov/specific-groups/people-at-higher-risk.html.
\bibitem{WHO} World Health Organization (2016) Ebola situation reports. https://apps.who.int/ebola/ebola-situation-reports (2016).
\bibitem{Ritchie2020} Roser M, Ritchie H (2020) ``Malaria''. https://ourworldindata.org/malaria.
\bibitem{Arcila2019} Nunes PCG, Daumas RP, S\'anchez-Arcila JC et al (2019) 30 years of fatal dengue cases in Brazil: a review. BMC Pub Heal 19:329.
\bibitem{Tabeau2001} Tabeau E, van den Berg Jeths A, Heathcote C (2001) Forecasting mortality in developed countries: insights from a statistical, demographic and epidemiological perspective. Springer Science \& Business Media.
\bibitem{Utsunomiya2020} Utsunomiya YT, Utsunomiya ATH, Torrecilha RBP et al (2020) Growth rate and acceleration analysis of the COVID-19 pandemic reveals the effect of public health measures in real time. Front Med 7:247.
\bibitem{percolation1} Sahini M (1994) Applications of percolation theory. Taylor \& Francis, London.
\bibitem{percolation2} Stauffer D, Aharony A (1994) Introduction to percolation theory. Taylor \& Francis, London.
\bibitem{superconductivity} Alexander S (1983) Superconductivity of networks. A percolation approach to the effects of disorder. Phys Rev B 27:1541.
\bibitem{semiconductors} Kaminski A, Das Sarma S (2002) Polaron percolation in diluted magnetic semiconductors. Phys Rev Lett 88:247202.
\bibitem{Li} Li D, Fu B, Wang Y et al (2015) Percolation transition in dynamical traffic network with evolving critical bottlenecks. PNAS 112(3):669-672.
\bibitem{epidemiology} Meyers LA (2007) Contact network epidemiology: bond percolation applied to infectious disease prediction and control. Bull Amer Math Soc 44:63-86.
\bibitem{Kotliar2004} Kotliar G, Vollhardt D (2004) Strongly correlated materials: insights from dynamical mean-field theory. Phys Today 57(3):53.
\bibitem{Metzner1989} Metzner W, Vollhardt D (1989) Correlated lattice fermions in $d = \infty$ dimensions. Phys Rev Lett 62:324.
\bibitem{Georges1994} Georges A, Kotliar G, Krauth W et al (1996) Dynamical mean-field theory of strongly correlated fermion systems and the limit of infinite dimensions. Rev     Mod Phys 68:13.
\bibitem{Riordan2006} Bollob\'as B, Riordan O (2006). Percolation. Cambridge University Press, Cambridge.
\bibitem{Choy} Choy TC (2015) Effective medium theory: principles and applications. Oxford University Press, Oxford.
\bibitem{Efros} Efros AL, Shklovskii BI (1976) Critical behaviour of conductivity and dielectric constant near the metal-non-metal transition threshold. Phys Stat Sol (b) 76:475-485.
\bibitem{vecteezy} Vectorized images available at https://www.vecteezy.com.
\bibitem{Mathematica} Weisstein EW ``Cayley Tree'' from mathworld - a Wolfram web resource. https://mathworld.wolfram.com/CayleyTree.html.
\bibitem{Florescu} Florescu I (2014) Probability and stochastic processes. John Wiley and Sons, Hoboken.
\bibitem{robotdance} Silva PJS, Pereira T, Nonato LG (2020) Robot dance: a city-wise automatic control of Covid-19 mitigation levels. medRxiv preprint. doi: https://doi.org/10.1101/2020.05.11.20098541.
\bibitem{Ralph} Baierlein R (1999) Thermal physics. Cambridge University Press, Cambridge.
\bibitem{Goltsev2008} Dorogovstev SN, Goltsev AV, Mendes JFF (2008) Critical phenomena in complex networks. Rev Mod Phys 80:1275.
\bibitem{Goltsev2015} Baxter GJ, Dorogovtsev SN, Lee KE et al (2015) Critical Dynamics of the $k$-core pruning process. Phys Rev X 5:031017.
\bibitem{Big-data}  Nagavci D, Hamiti M, Selimi B (2018) Review of prediction of disease trends using big data analytics. Int J Adv Comp Sci Appl 9(8):46-50.
\bibitem{Lee2018} Chae S, Kwon S, Lee D (2018) Predicting infectious disease using deep learning and big data. Int J Environ Res Public Health 15(8):1596.
\bibitem{Drake2016} Han BA, Drake JM (2016) Future directions in analytics for infectious disease intelligence. EMBO Rep 17:785-789.
\bibitem{Barber2011} Barber D (2012) Bayesian reasoning and machine learning. Cambridge University Press, New York.
\bibitem{Goodman2013} Foreman-Mackey D, Hogg DW, Lang D et al (2013) emcee: The MCMC hammer. Pub Astr Soc Pac 125(925):306-312.
\bibitem{Pedregosa2011} Pedregosa F, Varoquaux G, Gramfort A et al (2011) Scikit-learn: machine learning in Python. J Mach Learn Res 12:2825-2830.
\bibitem{Kou2015} Yang S, Sintillana M, Kou SC (2015) Accurate estimation of influenza epidemics using Google search data via ARGO. PNAS 112(47):14473-14478.
\bibitem{Santillana2020} Liu D, Clemente L, Poirier C et al (2020) A machine learning methodology for real-time forecasting of the 2019-2020 COVID-19 outbreak using Internet searches, news alerts, and estimates from mechanistic models. arXiv preprint. arXiv:2004.04019v1.
\bibitem{Morio} Morio J, Balesdent M (2016) Estimation of rare event probabilities in complex aerospace and other systems: a practical approach. Woodhead Publishing, Cambridge.
\bibitem{Heesterbeek2015} Heesterbeek H, Anderson RM, Andreasen V et al (2015) Modeling infectious disease dynamics in the complex landscape of global health. Science 347:6227.
\bibitem{Zhu2018} Zhu Y, Li C, Chen L et al (2018) A novel human coronavirus OC43 genotype detected in mainland China. Emerg Micr \& Inf 7(1):1-4.
\bibitem{Vespignani2020} Vespignani A, Tian H, Dye C et al (2020) Modelling COVID-19. Nat Rev Phys. doi:10.1038/s42254-020-0178-4.
\bibitem{Bedford2019} Bedford J, Farrar J, Ihekweazu C et al (2019) A new twenty-first century science for effective epidemic response. Nature 575:130–136.
\bibitem{Gbalo2019} Ebola Gbalo Research Group (2019) Responding to the Ebola virus disease outbreak in DR Congo: when will we learn from Sierra Leone? Lancet 393(10191):2647–2650.
\end{thebibliography}
% for author-year bibliography (bmc-mathphys or spbasic)
% a) write to bib file (bmc-mathphys only)
% @settings{label, options="nameyear"}
% b) uncomment next line
%\nocite{label}
\section*{Funding}
Not applicable.

\section*{List of abbreviations}
SIR, Susceptible, Immune, Recovered; MSEIR, Passively immune infants, Susceptible, Exposed, Infected, Recovered; MSEIRS, Passively immune infants, Susceptible, Exposed, Infected, Recovered, Susceptible; SEIR, Susceptible, Exposed, Infected, Recovered; SEIRS, Susceptible, Exposed, Infected, Recovered, Susceptible; SIRS, Susceptible, Infected, Recovered, Susceptible; SEI, Susceptible, Exposed, Infected; SEIS, Susceptible, Exposed, Infected, Susceptible; SI, Susceptible, Infected; SIS, Susceptible, Infected, Susceptible; WHO, World Health Organization; MG, Minas Gerais; SP, S\~ao Paulo; FD, Fermi-Dirac; FD-type, Fermi-Dirac-type; FL, Fermi-Liquid; CDC, Centers for Disease Control and Prevention; DOS, Density of States; DFT, Density Functional Theory; DMFT, Dynamical Mean-Field Theory; ARGO, AutoRegression with GOogle search data;

\section*{Availability of data and materials}

The datasets supporting the conclusions of this article are available in the repositories https://www.worldometers.info/coronavirus/; https://apps.who.int/ebola/ebola-situation-reports;
https://www.isglobal.org/en/ebola; https://condellpark.com/kd/sars.htm.

% or include bibliography directly:

%%%%%%%%%%%%%%%%%%%%%%%%%%%%%%%%%%%
%%                               %%
%% Figures                       %%
%%                               %%
%% NB: this is for captions and  %%
%% Titles. All graphics must be  %%
%% submitted separately and NOT  %%
%% included in the Tex document  %%
%%                               %%
%%%%%%%%%%%%%%%%%%%%%%%%%%%%%%%%%%%

%%
%% Do not use \listoffigures as most will included as separate files

\end{backmatter}
\end{document}